\documentclass{article}

\pdfoutput=1

\usepackage[linesnumbered,algo2e,lined,boxed,ruled,commentsnumbered]{algorithm2e}
\usepackage{microtype}
\usepackage{graphicx}
\usepackage{subfigure}
\usepackage{booktabs} 
\usepackage{enumitem}

\usepackage{hyperref}

\usepackage[accepted]{icml2024}

\usepackage{amsmath}
\usepackage{amssymb}
\usepackage{mathtools}
\usepackage{amsthm}
\usepackage{tablefootnote}

\allowdisplaybreaks

\usepackage[braket,qm]{qcircuit}
\usepackage{icml2024/nkj}

\usepackage[capitalize,noabbrev]{cleveref}

\graphicspath{{figures}{figures/visbo}}

\usepackage{listings}

\newif\ifdraft
\draftfalse

\usepackage[normalem]{ulem} 
\newcommand{\stkout}[1]{\ifmmode\text{\sout{\ensuremath{#1}}}\else\sout{#1}\fi}
\ifdraft
\newcommand{\added}[1]{\textcolor{blue}{#1}}
\newcommand{\deleted}[1]{\textcolor{purple}{\sout{#1}}}

\newcommand{\deletedfloat}[1]{\textcolor{blue}{#1}}
\else
\newcommand{\added}[1]{#1}
\newcommand{\deleted}[1]{}

\newcommand{\deletedfloat}[1]{}
\fi

\theoremstyle{plain}
\newtheorem{theorem}{Theorem}[section]

\newtheorem{corollary}[theorem]{Corollary}
\theoremstyle{definition}

\theoremstyle{remark}

\usepackage[textsize=tiny]{todonotes}

\newcommand{\methodname}{SubsCoRe} 
\newcommand{\totalshots}{2.5 \cdot 10^6} 

\icmltitlerunning{Adaptive Observation Cost Control for Variational Quantum Eigensolvers}

\begin{document}

\twocolumn[
\icmltitle{Adaptive Observation Cost Control for Variational Quantum Eigensolvers}



\icmlsetsymbol{equal}{*}

\begin{icmlauthorlist}
\icmlauthor{Christopher J.~Anders}{equal,affil1,affil2}
\icmlauthor{Kim A.~Nicoli}{equal,affil3,affil4}
\icmlauthor{Bingting Wu}{affil2}
\icmlauthor{Naima Elosegui}{affil1,affil2}
\icmlauthor{Samuele Pedrielli}{affil2}
\icmlauthor{Lena Funcke}{affil3,affil4}
\icmlauthor{Karl Jansen}{affil5}
\icmlauthor{Stefan Kühn}{affil5}
\icmlauthor{Shinichi Nakajima}{affil1,affil2,affil6}

\end{icmlauthorlist}

\icmlaffiliation{affil1}{Berlin Institute for the Foundations of Learning and Data (BIFOLD)}
\icmlaffiliation{affil2}{Technische Universit\"{a}t Berlin, Germany}
\icmlaffiliation{affil3}{Transdisciplinary Research Area (TRA) Matter, University of Bonn, Germany}
\icmlaffiliation{affil4}{Helmholtz Institute for Radiation and Nuclear Physics (HISKP), University of Bonn, Germany}
\icmlaffiliation{affil5}{Deutsches Elektronen-Synchrotron (DESY), Zeuthen, Germany}
\icmlaffiliation{affil6}{RIKEN Center for AIP, Japan}
\icmlcorrespondingauthor{}{anders/nakajima@tu-berlin.de}
\icmlcorrespondingauthor{}{knicoli@uni-bonn.de}

\icmlkeywords{Gaussian Processes, Quantum Computing, Variational Quantum Eigensolvers, VQE}

\vskip 0.3in
]



\printAffiliationsAndNotice{\icmlEqualContribution} 

\begin{abstract}
The objective to be minimized in the variational quantum eigensolver (VQE) has a restricted form, which allows a specialized sequential minimal optimization (SMO) that requires only a few observations in each iteration.  However, the SMO iteration is still costly due to the observation noise---one \emph{observation} at a point typically requires averaging over hundreds to thousands of repeated quantum \emph{measurement shots} for achieving a reasonable noise level.  
In this paper, we propose an adaptive cost control method,
named \emph{subspace in confident region} (\methodname{}), for SMO.
\methodname{} uses the Gaussian process (GP) surrogate, and requires it to have low uncertainty over the subspace being updated, so that optimization in each iteration is performed with guaranteed accuracy.  
The adaptive cost control is performed by first setting the required accuracy according to the progress of the optimization, and then choosing the minimum number of measurement shots and their distribution such that the required accuracy is satisfied.
We demonstrate that \methodname{} significantly improves the efficiency of SMO, and outperforms the state-of-the-art methods.
\end{abstract}

\section{Introduction}
\label{sec:Introduction}
\begin{figure}[t]
    \centering
    \includegraphics[width=0.460\textwidth]{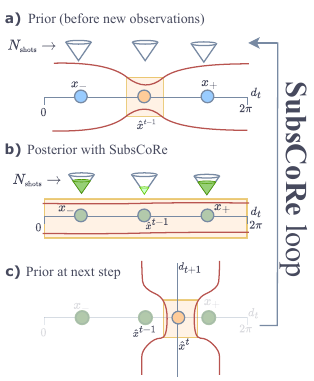}
    \vskip -1ex
    \caption{Illustration of \methodname{}. (a) After iteration $t-1$, the function value at the best point $\widehat{\bfx}^{t-1}$, which is in the CoRe (yellow rectangle) due to the previous observations, is predicted by the GP with sufficiently low uncertainty (red curves). Two equidistant shifts (blue points) at $\widehat{\bfx}^{t-1} \pm \frac{2\pi}{3} \bfe_{d_t}$ are chosen, yet not measured, along the line parallel to the $d_t$ axis at the current step. (b) \methodname{} finds the minimum number of measurement shots for measuring the points $\{\widehat{\bfx}^{t-1},\bfx_\pm\}$ such that the entire line will be included in the CoRe, i.e., the \textit{posterior uncertainty} at any point on the line is smaller than the required threshold. The three points are measured (green circles), with the middle point $\widehat{\bfx}^{t-1}$ typically requiring fewer shots because of its lower prior uncertainty. 
    (c) Using the trigonometric polynomial regression, the minimum $\widehat{\bfx}^{t}$ (orange point) along the line (where the GP mean function is fully identified by predicting \emph{any} three points) is computed along with the corresponding energy,
    thus becoming the starting point for the next iteration.
}
    \label{fig:cartoon_SubsCoRe}
    \vspace{-5mm}
\end{figure}
Introduced a decade ago by~\citet{Peruzzo2014}, the variational quantum eigensolver (VQE)~\citep{mcclean2016theory} has become a widespread hybrid quantum-classical algorithm for approximating the ground-state of a given Hamiltonian. It utilizes a quantum device to efficiently prepare a trial state in the form of a parameterized ansatz circuit and to measure the expectation value of the Hamiltonian, i.e., the energy, in the given ansatz. Subsequently, based on the measurement outcome, a classical optimizer is used to compute a new set of parameters that likely lowers the energy. Running the feedback loop between the quantum device and the classical optimizer until convergence, the parametric ansatz encodes an approximation for the ground-state of the Hamiltonian, and the final energy measurement provides an estimate for the ground-state energy.

VQEs may offer the potential to overcome computational challenges in many fields of science, including quantum chemistry, biology, and material science~\cite{kandala2017hardware,grimsley2019adaptive,bauer2020quantum,fedorov2021towards,PhysRevResearch.2.043140},
where accurately identifying the ground-state energy of complex molecules or materials is intractable for classical computers. By leveraging the principles of quantum mechanics, VQEs may efficiently solve such problems, enabling advancements in drug discovery, materials design, and other fields~\citep{cao2018potential}.
Furthermore, applications of VQEs to other computationally challenging tasks, e.g., high-energy physics calculations~\cite{Humble2022,DiMeglio:2023nsa} including lattice field theory simulations~\cite{Funcke:2023jbq}, have been recently investigated. We refer readers to~\citet{tilly2022variational} for a comprehensive review.

As is typical for any type of quantum computing protocols, 
VQEs suffer from two types of noise---hardware noise and shot noise.
The hardware noise is caused by the imperfect quantum circuit that systematically or stochastically distorts the observations, while the shot noise is due to the intrinsic (and therefore unavoidable) stochastic nature of quantum computing. 
Throughout the paper, we do not consider the hardware noise%
\footnote{
Recent progress in quantum hardware development has substantially reduced the hardware noise, see, e.g.,~\citet{bluvstein2023logical}.
}
and focus on the \textit{shot noise}, i.e., statistical uncertainties due to a finite number of repeated measurements (shots) to estimate expected values of observables on the quantum computer.
We treat the shot noise as i.i.d. \emph{observation noise} when regression is applied to the measurements.

For quantum circuits that consist of parametrized rotation gates and non-parametric unitary gates (e.g., the Hadamard gate and the Controlled-Z gate), it is known that the objective function in VQE is a low-order trigonometric polynomial along each axis, and therefore can be fit exactly by a linear model with low degrees of freedom~\citep{nakanishi20}.  Based on this property,  \citet{nakanishi20} proposed an efficient sequential minimal optimization (SMO) \citep{Platt1998SequentialMO}---often referred to as Nakanishi-Fuji-Todo (NFT)---which requires only a few (two in the most typical case) observations in each iteration of the 1D subspace optimization.  \citet{NEURIPS:Nicoli+:2023} introduced the corresponding VQE kernel, and enhanced NFT with Gaussian processes (GP).  Specifically, they applied Bayesian optimization (BO) to find the best locations to be observed in each SMO iteration.  To make use of the property that only a few points determine the complete functional form in the 1D subspace, they introduced the notion of confident region (CoRe)---the region on which the uncertainty of the GP prediction is lower than a threshold---and proposed the expected maximum improvement over CoRe (EMICoRe) acquisition function to evaluate how much the CoRe extends and how much improvement is expected within the CoRe after new points are observed.  BO with EMICoRe successfully enhanced NFT, proving the usefulness of GP for VQE~\citep{NEURIPS:Nicoli+:2023}.

In this paper, we enhance NFT in an orthogonal way---introducing adaptive observation cost control.  
Due to the shot noise, each \emph{observation}, i.e., each evaluation of the cost function value at a single point, consists of hundreds to thousands of \emph{measurement shots}---a repeated quantum computation from the initial state preparation to the final quantum state observation, often also referred to as \textit{readouts}. The computational cost on the quantum computer is largely dominated by this quantum process and should, therefore, be measured by the total number of shots.  This implies that one can improve the computational efficiency by adapting the number of shots according to the progress of optimization, e.g., saving shots in the early phase where the best value rapidly decreases, and using more shots in the converging phase where accurate prediction guides the optimizer towards the very bottom of the objective, i.e., the ground-state. Such strategies adapting the number of shots have been proposed and applied to gradient-based methods. For example, \citet{SGLBO2022} proposed the stochastic gradient line BO (SGLBO) approach, where the number of shots is set according to the norm of the estimated gradient, and showed that SGLBO outperforms NFT with a fixed number of shots.
An adaptive-shots strategy for NFT---where the gradient is not estimated---is yet to be established and is the primary goal of this paper.

Our approach, named \emph{subspace in confident region} (\methodname{}), distributes the minimum number of shots in each SMO iteration so that the uncertainty in the updated subspace is lower than a threshold. In other words, we search for the minimum number of shots, i.e., minimize the observation cost, such that the optimized subspace is a subset of the CoRe (see \Cref{fig:cartoon_SubsCoRe}). 
Although identifying the optimal distribution of thousands of shots seems intractable, our theory implies that we do not need to distribute the shots thinly over the subspace.  More specifically, observing $1+2V_d$ equidistant points, where $V_d$ is the number of gates parameterized by the $d$-th variable,
with equally divided shots results in a min-max optimal uniform posterior uncertainty.
Based on our theory, we propose a few variants of \methodname{} and show their state-of-the-art performance.

\paragraph{Related Work}
Some of the recently proposed optimization techniques for VQEs
rely on special properties of the VQE objective, i.e., the parameter shift rule (PSR)~\citep{Mitarai18,schuld2019evaluating} and the low-order trigonometric
polynomial form along each axis~\citep{nakanishi20}, which have been shown to be equivalent \citep{NEURIPS:Nicoli+:2023}.
The PSR is used for efficient gradient estimation~\citep{Mitarai18}, while the trigonometric polynomial form allows for efficient SMO, i.e., NFT \citep{nakanishi20}.
Vanilla BO with a GP surrogate was also applied to VQE~\citep{iannelli2021noisy,SGLBO2022,NEURIPS:Nicoli+:2023}.
\citet{SGLBO2022} combined stochastic gradient descent and 1D BO with an adaptive number of measurement shots, which outperforms NFT with a \textit{fixed} number of shots.  \citet{NEURIPS:Nicoli+:2023} introduced the VQE kernel and the EMICoRe acquisition function, and enhanced NFT with BO.
Although the usefulness of GP for VQE has already been reported, to the best of our knowledge, no prior work has used it for observation cost control.
\added{In the more general BO context, works dealing with heteroscedastic noise by incorporating the variance in the acquisition function for BO and active learning have also been proposed~\cite{doi:10.1080/00401706.2018.1469433,NEURIPS2021_8f97d1d7}.}

\added{\paragraph{Difference between EMICoRe and \methodname{}}
\label{sec:differenceAlgorithms}
All approaches analyzed in our numerical experiments, i.e, the NFT, EMICoRe, SGLBO and the newly proposed \methodname{}, differ substantially from each other. While the difference between \methodname{} and SGLBO is apparent, here we elaborate on the crucial differences between EMICoRe and \methodname{}. A key feature of both \methodname{} and EMICoRe is that both approaches rely on the notion of confident region (CoRe) and on the VQE kernel introduced by~\citet{NEURIPS:Nicoli+:2023}. However, besides this, the objective and the optimization of both approaches substantially differ. EMICoRe uses a special acquisition function (based on the CoRe) to select the best points to measure at the next iteration of SMO, and the observations are performed at a fixed number of shots (repeated measurements on a quantum computer).  Namely, it tries to minimize the number of SMO steps required to converge with a fixed number of shots for each observed point. This approach is fundamentally different from \methodname{}, which focuses on minimizing the total quantum computing budget, i.e., the total number of shots, regardless of the number of classical optimization steps. We stress that in VQEs the crucial bottleneck is the number of operations performed on the quantum device,
and, therefore, minimizing the total number of shots as the ``observation cost'' is more appropriate than minimizing the SMO steps required for convergence.
In addition, \methodname{} relies on the min-max optimality of the equidistant measurements under the ``subspace in the CoRe'' requirement, and thus observes fixed points in each SMO step without using acquisition functions.
\methodname{} adapts the number of shots
based on how the CoRe extends after the measurements, and, therefore, the necessary number of shots is inherently tied to the confidence of the GP.  More specifically, our algorithm chooses the number of measurement shots at each observed point such that the posterior variance is within a given threshold, which is adapted during the optimization. To summarize, both the methodology and the objective of EMICoRe and \methodname{} are fundamentally different, despite both relying on the notion of CoRe. Furthermore, we note that both approaches are not mutually exclusive and can be combined. We defer such a study to future work.}

\section{Background}
\label{sec:Background}

\subsection{Gaussian Process (GP) Regression with Heteroscedastic Observation Noise}
\label{sec:Pre.GP}

Assume that we aim to learn an unknown function $f^*(\cdot): \mcX \mapsto \bbR$ from the training data
$\bfX = ({\bfx}_1, \ldots, {\bfx}_N) \in \mcX^{N}, \bfy = ( y_1, \ldots, y_{N})^\T \in \bbR^{N}, \bfsigma = (\sigma_1^2, \ldots, \sigma_N^2) \in \mathbb{R}_{++}^N$
that fulfills
\begin{align}
y_n &= f^*(\bfx_n) + \varepsilon_n, & \varepsilon_n &\sim  \mcN_1(y_n; 0, \sigma_n^2),
\label{eq:RegressionModel}
\end{align}
where $\mcN_D(\cdot; \bfmu, \bfSigma)$ denotes the $D$-dimensional Gaussian distribution with mean $\bfmu$ and covariance $\bfSigma$.
With the Gaussian process (GP) prior
\begin{align}
p(f(\cdot))
&= \mathrm{GP} (f(\cdot); 0(\cdot), k(\cdot, \cdot)),
\label{eq:GPPrior}
\end{align}
where $0(\cdot)$ and $k(\cdot, \cdot)$ are the prior zero-mean and the kernel (covariance) functions, respectively,
the posterior distribution of the function values $\bff' = (f(\bfx'_1), \ldots, f(\bfx'_M))^\T \in \bbR^M$ at arbitrary test points $\bfX' = ({\bfx'}_1, \ldots, {\bfx'}_M) \in \mcX^{M}$ is given as
\begin{align}
p({\bff'} | \bfX, \bfy)  & =  \mcN_{M}({\bff'}; \bfmu'_{[ \bfX, \bfy, \bfsigma]}, \bfS'_{[ \bfX, \bfsigma]}), \quad \mbox{ where }
   \label{eq:GPPosterior}\\
\bfmu'_{[ \bfX, \bfy, \bfsigma]}
   &= {\bfK}'^{\T} \left(\bfK + \bfDiag(\bfsigma) \right)^{-1} \bfy,
 \label{eq:GPPosteriorMean}\\
\bfS'_{[ \bfX, \bfsigma]}
   &= {\bfK}'' - 
   {\bfK'}^{\T} \left(\bfK + \bfDiag(\bfsigma) \right)^{-1} {{\bfK}'}
   \label{eq:GPPosteriorVar}
\end{align}
 \citep{book:Rasmussen+Williams:2006}.
Here $\bfDiag(\bfv)$ is the diagonal matrix with $\bfv$ specifying the diagonal entries,
and $\bfK = k(\bfX, \bfX) \in \bbR^{N \times N}, {\bfK}'  = {k}(\bfX, \bfX') \in \bbR^{N \times M}$, and $  {\bfK}'' ={k}(\bfX', \bfX')  \in \bbR^{M \times M}$ are the train, train-test, and test kernel matrices, respectively,
where $k(\bfX, \bfX')$ denotes the kernel matrix evaluated at each column of $\bfX$ and $\bfX'$ such that $(k(\bfX, \bfX'))_{n, m} = k(\bfx_n, \bfx'_m)$.  We also denote the posterior as $p({f}(\cdot) | \bfX, \bfy)   =  \mathrm{GP} (f(\cdot); \mu_{[ \bfX, \bfy, \bfsigma]}(\cdot), s_{[ \bfX, \bfsigma]}(\cdot, \cdot))$ with the posterior mean $\mu_{[ \bfX, \bfy, \bfsigma]}(\cdot)$ and covariance $s_{[ \bfX, \bfsigma]}(\cdot, \cdot)$ functions, e.g., $\mu_{[ \bfX, \bfy, \bfsigma]}(\bfx') = \bfmu_{[ \bfX, \bfy, \bfsigma]}' \in \mathbb{R}$ and $s_{[ \bfX, \bfsigma]}(\bfx', \bfx') = \bfS_{[ \bfX, \bfsigma]}'  \in \mathbb{R}_{++}$ for a single test point $\bfX' = (\bfx')$.

\subsection{Variational Quantum Eigensolvers}
\label{sec:B.VQE}

The VQE~\citep{Peruzzo2014,mcclean2016theory} is a hybrid quantum-classical computing protocol for estimating the ground-state energy of a given quantum Hamiltonian for a $Q$-qubit system.
A quantum computer is used to prepare a trial state $\vert\psi_{0}\rangle$ transformed via a parametric quantum circuit $G(\bfx)$, which depends on $D$ angular parameters $\bfx \in \mcX = [0, 2 \pi)^D$. 
The final state $\vert\psi_{\bfx}\rangle$ is thus generated by applying $D' (\geq D)$ \emph{quantum gate operations}, $G(\bfx) = G_{D'} \circ\cdots \circ G_1$, to the initial trial state $\vert{\psi_0}\rangle$, i.e.,  $\vert\psi_{\bfx} \rangle = G(\bfx) \vert\psi_0\rangle$. 
All gates $\{G_{d'}\}_{d'=1}^{D'}$ are unitary operators, parameterized by at most one variable $x_d$. 
Let $d(d'): \{1, \ldots, D'\} \mapsto \{1, \ldots, D\}$ be the mapping specifying which one of the variables $\{x_d\}$ parameterizes the $d'$-th gate.
We consider parametric gates of the form $G_{d'}(x) = U_{d'} (x_{d(d')}) = \exp \left( -i x_{d(d')} P_{d'}/2 \right)$, where $P_{d'}$ is an arbitrary sequence of the Pauli operators $ \{\mathbf{1}_q,\,\sigma_q^X, \sigma_q^Y, \sigma_q^Z\}_{q=1}^Q$ acting on each qubit at most once.
This general structure covers both single-qubit gates, such as $R_{X}(x) = \exp{\left(-i\theta \sigma_q^X \right)}$, and entangling gates acting on multiple qubits simultaneously, such as $R_{XX}(x) = \exp{\left(-i x \sigma_{q_1}^X \circ \sigma_{q_2}^X \right)}$ and $R_{ZZ}(x) = \exp{\left(-i x \sigma_{q_1}^Z \circ \sigma_{q_2}^Z\right)}$ for $q_1 \ne q_2$, commonly realized in trapped-ion quantum hardware setups~\citep{TrappedIon2,TrappedIon}. 

The quantum computer is used to evaluate the energy of the resulting quantum state $\ket{\psi_{\bfx}}$
by observing
\begin{align}
y &= f^*(\bfx) + \varepsilon,
\qquad \mbox{ where }
\notag\\
f^*(\bfx)
&=
\langle{\psi_{\bfx}}\vert  H  \vert{\psi_{\bfx}}\rangle
=
\langle{\psi_0}\vert G(\bfx)^\dagger H G(\bfx) \vert{\psi_0}\rangle.
\label{eq:VQEObjective}
\end{align}
Here $\dagger$ denotes the Hermitian conjugate. 
For each observation, multiple measurement shots, denoted by $N_\mathrm{shots}$, are acquired to suppress the variance $ \sigma^{*2} (N_\mathrm{shots})$ of the noise $\varepsilon$ in evaluating the expectation value of the Hamiltonian, i.e., the energy of the quantum system.%
\footnote{
When the Hamiltonian consists of $N_\mathrm{{og}}$ groups of non-commuting operators, each of which needs to be measured separately, $N_\mathrm{shots}$ denotes the number of shots \emph{per operator group}. Therefore, the number of shots \emph{per observation} is $N_\mathrm{{og}}\times N_\mathrm{shots}$.
Throughout the paper, we use the total number of shots per operator group, i.e., the cumulative sum of $N_\mathrm{{shots}}$ over all observations, to evaluate the observation cost.
}
Since the observation $y$ is the sum of many random variables, it approximately follows the Gaussian distribution, according to the central limit theorem. The Gaussian likelihood~\eqref{eq:RegressionModel} therefore approximates the observation $y$ well if 
$\sigma^2 \approx  \sigma^{*2} (N_\mathrm{shots})$.
Using the noisy estimates of $f^*(\bfx)$ obtained from the quantum computer, a protocol running on a classical computer is used to solve the following minimization problem:
\begin{align}
\textstyle
\min_{\bfx \in [0, 2 \pi)^D} f^*(\bfx),
\label{eq:VQEOptimization}
\end{align}
thus finding the minimizer $\widehat{\bfx}$, i.e., the optimal parameters for the 
(rotational) quantum gates.

Let $V_d$ be the number of gates parameterized by $x_d$, i.e., $ V_d= |\{d' \in \{1,\dots D'\}; d= d(d')\} |$, and
\begin{align}
    \bfpsi_{\gamma}(\theta) 
    &= (\gamma, \sqrt{2} \cos \theta, \sqrt{2} \cos2 \theta, \ldots, \sqrt{2} \cos V_d \theta,
    \notag\\
    & \hspace{-5mm}
    \sqrt{2} \sin \theta, \sqrt{2} \sin 2 \theta, \ldots, \sqrt{2} \sin V_d \theta)^\T \in \mathbb{R}^{1 + 2V_d}
    \label{eq:FourierBasis}
\end{align}    
be the (1-dimensional) $V_d$-th order Fourier basis for arbitrary $\gamma > 0$. 
\citet{nakanishi20} found that 
the VQE objective function $f^*(\cdot)$ in Eq.~\eqref{eq:VQEObjective} with any\footnote{Any circuit consisting of parametrized rotation gates and non-parametric unitary gates as stated in the introduction.} $G(\cdot)$, $H$, and $\ket{\psi_0}$ can be exactly expressed as
\begin{align}
f^*(\bfx) =  \bfb^\T  \mathbf{vec} \left( \otimes_{d=1}^D 
\bfpsi_{\gamma}(x_d)
\right)
\label{eq:TrigonometricPolynomial}
\end{align}
for some $\bfb  \in \textstyle  \mathbb{R}^{\prod_{d=1}^D(1 + 2V_d)}$,
where 
$\otimes$ and $\mathbf{vec} (\cdot)$ denote the tensor product and the vectorization operator for a tensor, respectively.
Based on this property,
the NFT~\citep{nakanishi20} method performs SMO~\citep{Platt1998SequentialMO}, where the optimum in the 1D subspace in each iteration is analytically estimated from only $1+2V_d$ observations.

\paragraph{Nakanishi-Fuji-Todo (NFT) Algorithm}
Let $\{\bfe_d\}_{d=1}^{D}$ be the standard basis.
NFT is initialized with a random point $\widehat{\bfx}^{0}$ with a first observation $\widehat{y}^0 = f^*(\widehat{\bfx}^0) + \varepsilon_0$, and iterates the following procedure: for each iteration step $t$, 
\begin{enumerate}
\itemsep0em 
    \item Select an axis $d \in \{1, \ldots, D\}$ sequentially and observe the objective $\bfy \in \mathbb{R}^{2V_d}$ at $2V_d$ points 
$\bfX = (\bfx_1, \ldots, \bfx_{2V_d}) =  \{ \widehat{\bfx}^{t-1} +  \alpha_w  \bfe_{{d}}  \}_{w=1}^{2V_d} \in \mathbb{R}^{D \times 2V_d}$ along the axis $d$.%
\footnote{
With slight abuse of notation, we use the set notation to specify the column vectors of a matrix, i.e., $(\bfx_1, \ldots, \bfx_N) = \{\bfx_n\}_{n=1}^N$.
}
Here $\bfalpha \in [0, 2 \pi)^{2V_d}$ is such that $\alpha_w \ne 0,\, \alpha_{w'} \ne \alpha_w$, for all $w$ and $w' \ne w$.

    \item Apply the 1D trigonometric polynomial regression $\widetilde{f}(\theta) =  \widetilde{\bfb}^\T \bfpsi_{1} (\theta)$ to the $2V_d$ new observations $\bfy$, together with the previous best estimated score $\widehat{y}^{t-1}$, and analytically compute the new optimum $\widehat{\bfx}^t = \widehat{\bfx}^{t-1} +  \widehat{\theta} \bfe_{{d}}$, where $\widehat{\theta} =\argmin_{\theta }\widetilde{f}(\theta)$. 
    \item Update the best score by $ \widehat{y}^t =  \widetilde{f}(\widehat{\theta})$.
\end{enumerate}
Note that if the observation noise is negligible, i.e., $y \approx f^*(\bfx)$, 
each step of NFT reaches the global optimum in the 1D subspace along the chosen axis $d$ for any choice of $\bfalpha$, and thus performs SMO exactly. Otherwise, errors can be accumulated in the best score $\widehat{y}^t$, and therefore an additional measurement may need to be performed at $\widehat{\bfx}^t$ after a certain iteration interval.

Inspired by the trigonometric polynomial form
\eqref{eq:TrigonometricPolynomial},
\citet{NEURIPS:Nicoli+:2023} used the corresponding VQE kernel
\begin{align}
k_{\gamma} (\bfx, \bfx')
&= \textstyle
\sigma_0^2
\prod_{d=1}^D \left(\frac{ \gamma^{2} + 2\sum_{v=1}^{V_d}  \cos \left( v(x_d - x_d')  \right)}{\gamma^{2} + 2 V_d }\right),
\label{eq:VQEKernelTied}
\end{align}
which is decomposed as $k_{\gamma} (\bfx, \bfx') = \bfphi_{\gamma}(\bfx)^\T \bfphi_\gamma(\bfx')$ for $\bfphi_{\gamma}(\bfx) = 
 \frac{\sigma_0 }
{ \left( \gamma^{2} + 2V_d  \right)^{D/2}} \mathbf{vec} \left( \otimes_{d=1}^D 
\bfpsi_{\gamma}(x_d)
\right)$,
for GP regression, and enhanced NFT with BO, using the notion of CoRe
\begin{align}
\mcZ_{[\bfX, \bfsigma]}(\kappa^2) & = \left\{\bfx \in \mcX; s_{[\bfX, \bfsigma]} (\bfx, \bfx) \leq \kappa^2 \right\},
\label{eq:LowUncertaintySet}
\end{align}
i.e., the region in which the uncertainty of the GP prediction is lower than a threshold $\kappa$. Specifically, \citet{NEURIPS:Nicoli+:2023} introduced the EMICoRe acquisition function to find the best observation points, i.e., $\bfalpha$ in Step 1 of NFT, such that the maximum expected improvement within CoRe is maximized.
Note that the kernel parameter $\gamma^2$ controls the smoothness of the function, i.e., suppressing the interaction terms when $\gamma^2 > 1$. When $\gamma^2 = 1$, the Fourier basis \eqref{eq:FourierBasis} is orthonormal, and 
the VQE kernel \eqref{eq:VQEKernelTied} is proportional to the product of Dirichlet kernels \citep{Rudin1962}.

\begin{figure*}[t]
    \centering
    \includegraphics[height=7ex,]{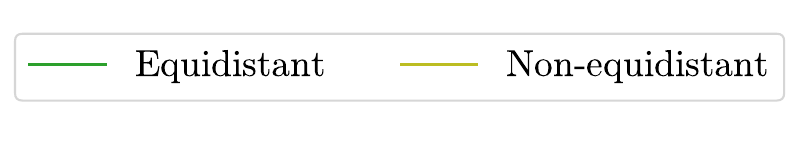}\\
    \includegraphics[width=0.4\textwidth]{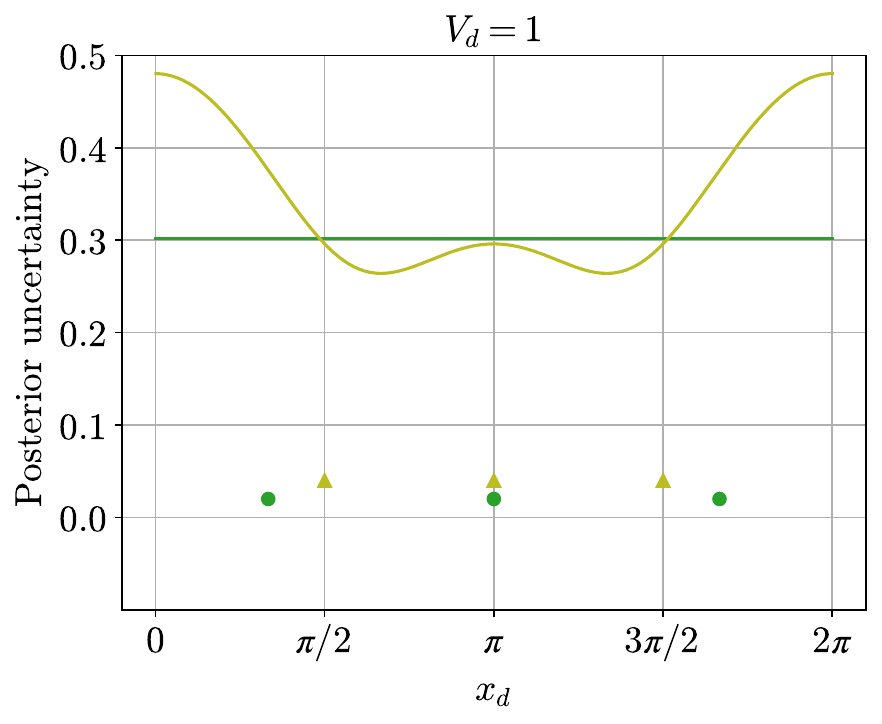}
    \hspace{10mm}
    \includegraphics[width=0.4\textwidth]{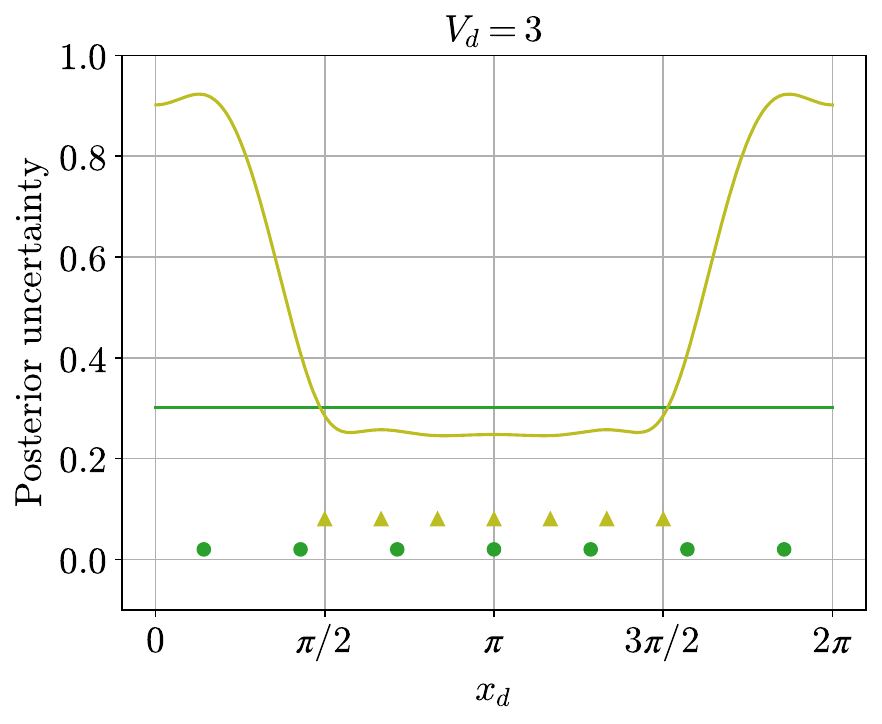}
    \centering\vskip -2ex
    \vskip -1ex
    \caption{Uncertainty of GP with the VQE kernel trained on equidistant \added{(darker green, round markers)} and non-equidistant \added{(olive, triangle markers)} observation points. The left and right plots show the $V_d=1$ and $V_d=3$ cases, respectively. \added{The green and olive solid lines refer to the GP posterior uncertainty obtained after observing $1 + 2 V_d$ points with an equidistant and non-equidistant spacing. 
    For instance, in the $V_d=1$ case,
    the observations along a one-dimensional subspace (parallel to the $d$-axis) are performed at 
    $x_d=\pi - \alpha, \pi, \pi + \alpha$
     with $\alpha$ being $\frac{\pi}{2}$ (olive) and $\frac{2\pi}{3}$ (green) in the non-equidistant and equidistant cases, respectively.}
    Following our theory, observing $1 + 2V_d$ equidistant points leads to uniform posterior uncertainty.
    }
    \label{fig:TheoreticalUncertainty}
\end{figure*}

\section{Subspace in Confident Region (\methodname{})}
\label{sec:ProposedMethod}

We propose an  adaptive observation cost control method, named \emph{subspace in confident region} (\methodname{}), for NFT, where the notion of CoRe \citep{NEURIPS:Nicoli+:2023} for GP plays an essential role.  Specifically, in each NFT iteration $t$, \methodname{} uses the minimum total number of measurement shots and distributes them optimally, so that the subspace optimization is performed with the required accuracy $\kappa_t^2$.  This is guaranteed by requiring that the entire subspace
$
\mcS_d(\widehat{\bfx}) = \{\widehat{\bfx} + {\alpha}' \bfe_d; {\alpha}' \in [0, 2\pi)\}
$
to be updated is a subset of the CoRe, i.e.,
\begin{align}
    \mcS_d(\widehat{\bfx}) \subseteq
 \mcZ_{[(\bfX, \breve{\bfX}),(\bfsigma, \breve{\bfsigma})]} (\kappa_t^2),
 \label{eq:SubscoreRequirement}
 \end{align}
 when the GP is trained on the augmented data $(\bfX, \breve{\bfX}),  (\bfsigma, \breve{\bfsigma})$ with new observation points $\breve{\bfX}, \breve{\bfsigma}$.
 Let $\overline{N}_{\mathrm{shot}}^t$ be the total number of shots we use in the $t$-th iteration.
 In one extreme, we can allocate each single shot to an arbitrary position, i.e., choose $\bfalpha \in [0, 2\pi)^{\overline{N}_{\mathrm{shot}}^t}$ and set $\breve{\bfX} = \{\widehat{\bfx} + \alpha_w \bfe_d\}_{w=1}^{\overline{N}_{\mathrm{shot}}^t} \in \mathbb{R}^{D \times \overline{N}_{\mathrm{shot}}^t}$ and $\breve{\bfsigma} = \sigma^{*2} (1) \bfone_{\overline{N}_{\mathrm{shot}}^t}$, where 
$\sigma^{*2} (1) = \overline{\sigma}^{*2} $ is the observation noise variance for single-shot measurements, and $\bfone_N$ denotes the all-one vector in the $N$-dimensional space. \added{Here, we note that the observation noise variance is defined as a function of $N_{\mathrm{shots}}$ to be $$\textstyle \sigma^{*2}(N_{\mathrm{shots}})=\frac{\overline{\sigma}^{*2}}{N_{\mathrm{shots}}}. $$ 
}  The optimum number of shots and their distribution are given by solving 
\begin{align}
\min_{\overline{N}_{\mathrm{shot}}, \bfalpha \in [0, 2\pi)^{\overline{N}_{\mathrm{shot}}^t}}
\!\!\!\!\! \!
\overline{N}_{\mathrm{shot}} \mbox { s.t. } 
\mcS_d(\widehat{\bfx}) \!\subseteq\!
 \mcZ_{[(\bfX, \breve{\bfX} (\bfalpha)), (\bfsigma, \breve{\bfsigma})]} (\kappa_t^2).
\label{eq:OptimalSubsCoReProblem}
\end{align}

The following theory implies that we do not need to solve such a huge $(\overline{N}_{\mathrm{shot}} \sim 10^3)$-dimensional problem, and guides us to simple approximate solutions (the proof is given in \Cref{sec:A.Proofs}):
\begin{theorem}
\label{thrm:UniformUncertainty}
Assume that, for an arbitrary $\widehat{\bfx} \in [0, 2\pi)^D$, we observed the function values $\bfy  \in \mathbb{R}^{1 + 2V_d}$ at the equidistant points 
$\bfX = \{\widehat{\bfx}  + \alpha_w \bfe_d; \alpha_w = \frac{2w}{1 + 2V_d } \pi \}_{w=0}^{2V_d}$   along the $d$-th axis 
 with the same observation noise variance $\sigma^2$.
  Then, the posterior variance of GP with the VQE kernel at the test point $\bfx' = \widehat{\bfx} + \alpha' \bfe_d$ for arbitrary $\alpha' \in [0, 2\pi)$ is 
\begin{align}
s_{\bfX}(\bfx'\!, \bfx')
& \!= \!\textstyle   \frac{  \sigma^2 \left( \!(\gamma^2 + 2V_d)^2 \frac{\sigma^{2}}{\sigma_0^2}  + (1+2V_d)^2\gamma^2\! \right)  }{\left(\!(\gamma^2 + 2V_d) \frac{\sigma^{2}}{\sigma_0^2} + 1 + 2V_d \!\right)\left(\!( \gamma^2 + 2V_d)\frac{\sigma^{2}}{\sigma_0^2}  +(1 + 2V_d) \gamma^2\! \right) }.
\label{eq:UniformPredictiveUnvertainty}
\end{align}
\end{theorem}
This theorem assumes that we do not use the previous observed points $\bfX$, and only use new $1 + 2V_d$ observed points (including the point at $\widehat{\bfx}$ or $\alpha_0 = 0$).

Notably, the posterior variance does not depend on $\alpha'$, and is therefore uniform along the 1D subspace.  Figure~\ref{fig:TheoreticalUncertainty} depicts the GP uncertainty with equidistant and non-equidistant observed points for $V_d = 1$ (left) and $V_d = 3$ (right).  The left plot shows that, in terms of the worst uncertainty within the subspace, the optimal choice%
\footnote{
\added{Here the ``optimality'' is with respect to the minimization \eqref{eq:OptimalSubsCoReProblem} of the total number of shots that makes the entire updated subspace within the CoRe.
For a fixed observation budget, the equidistant observed points are proved to give the uniform uncertainty, which is min-max optimal, i.e., it minimizes the maximum uncertainty. The minimum observation cost can be obtained by adjusting the total number of shots, while keeping the optimized observed points, so that the uniform uncertainty~\eqref{eq:UniformPredictiveUnvertainty} matches the required accuracy $\kappa$.  Note that this does not necessarily mean that the equidistant measurements are optimal for the overall VQE performance. We defer further analysis and a full ablation study relating the optimality to the choice of the shift to future work, e.g., by combining EMICoRe and \methodname{}.}
}
for the $V_d=1$ case is not $\bfalpha = (0, \pi/2, 3\pi/2)^\T =(0, \pi/2, -\pi/2)^\T$---the setting used in the original NFT \citep{nakanishi20} ---but the equidistant
$\bfalpha = (0, 2\pi/3, 4\pi/3)^\T = (0, 2\pi/3, -2\pi/3)^\T$---the setting used for the baseline NFT in \citet{NEURIPS:Nicoli+:2023}. 

One might have naturally expected the uniform uncertainty at the observed equidistant points $\{\widehat{\bfx}  + \frac{2w}{1 + 2V_d } \pi \bfe_d\}_{w = 0}^{2V_d}$ from the viewpoint of the Fourier analysis: GP with the VQE kernel is equivalent to Bayesian linear regression with the Fourier basis \eqref{eq:FourierBasis}, and therefore the equidistant observations result in a regularized version of the discrete Fourier transform (see \Cref{sec:A.RelationToFourier}).
However, the uniform uncertainty between the equidistant points is not trivially expected, because the posterior variance \eqref{eq:GPPosteriorVar} has a quadratic form of $\bfK'$, which can contain $2V_d$-th order components (i.e., twice higher frequency than the highest frequency of the trigonometric polynomial~\eqref{eq:TrigonometricPolynomial}), as observed in the non-equidistant case in Figure~\ref{fig:TheoreticalUncertainty}.
One might have expected that the middle points between the neighboring observed points would have higher uncertainty, a belief disproved by~\Cref{thrm:UniformUncertainty}. 

An important practical implication of \Cref{thrm:UniformUncertainty} is that we can achieve uniform uncertainty---implying the min-max optimality---by observing $1 + 2V_d$ equidistant points with $\overline{N}_{\mathrm{shot}}^t/(1 + 2V_d)$ shots each, which drastically simplifies the problem \eqref{eq:OptimalSubsCoReProblem}.
Furthermore, it is crucial for GP with cubic complexity not to use too many training samples with large observation noise, although the classical computation cost is not counted in this paper.
The following corollary implies an even simpler alternative.
\begin{corollary}   
\label{thrm:UniformUncertaintyUpperBound}
The predictive uncertainty is upper-bounded by the observation noise variance, i.e.,
\begin{align}
s_{\bfX}(\bfx', \bfx') < \sigma^2.
\notag
\end{align}
\end{corollary}

Based on the theory above, we propose variants of \methodname{}, all of which observe the fixed $1 +2V_d$ equidistant points 
$\breve{\bfX} =   \{ \widehat{\bfx}^{t-1} +  \alpha_w  \bfe_{{d}} ; \alpha_w = \frac{2w}{1 + 2V_d } \pi  \}_{w=0}^{2V_d}$
with adapted observation noise variance $\breve{\bfsigma} \in \mathbb{R}_{++}^{1 + 2V_d}$.  Ignoring the rounding error, the total number of shots is proportional to $\overline{N}_{\mathrm{shot}} \propto (\|\breve{\bfsigma}\|_1)^{-1}$.  Therefore, under our fixed choice of $\breve{\bfX}$, the \methodname{} problem \eqref{eq:OptimalSubsCoReProblem} reduces to
\begin{align}
\max_{\breve{\bfsigma} \in \mathbb{R}_{++}^{1 + 2V_d}}
\|\breve{\bfsigma}\|_1 \mbox { s.t. } 
\mcS_d(\widehat{\bfx}) \subseteq
 \mcZ_{[(\bfX, \breve{\bfX}), (\bfsigma, \breve{\bfsigma})]} (\kappa_t^2).
\label{eq:OptimalSubsCoReProblemReduced}
\end{align}
The following variants give approximate feasible solutions.
\paragraph{\methodname{}-Bound}\label{alg.SubBound}
We set 
$\breve{\bfsigma} = \kappa_t^2 \bfone_{1 + 2V_d}$, which 
satisfies the \methodname{} requirement, i.e., 
the constraint in Eq.~\eqref{eq:OptimalSubsCoReProblemReduced},
according to 
\Cref{thrm:UniformUncertaintyUpperBound}.


\paragraph{\methodname{}\added{-Center}}\label{alg.SubCenter}
We set $\breve{\bfsigma} = (\breve{\sigma}_0^2, \breve{\sigma}^2, \ldots, \breve{\sigma}^2 )^{\T}$, i.e., we tie the entries except for the variance $\breve{\sigma}_0^2$ at the previous best point $\widehat{\bfx}^{t-1}$, and solve the \methodname{} problem \eqref{eq:OptimalSubsCoReProblemReduced} by a 2D search with respect to $\breve{\sigma}_0^2$ and $\breve{\sigma}^2$.

\begin{figure*}[t]
    \centering
    \includegraphics[height=4.5ex]{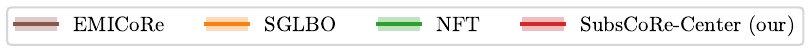}\\
    \includegraphics[width=0.49\textwidth]{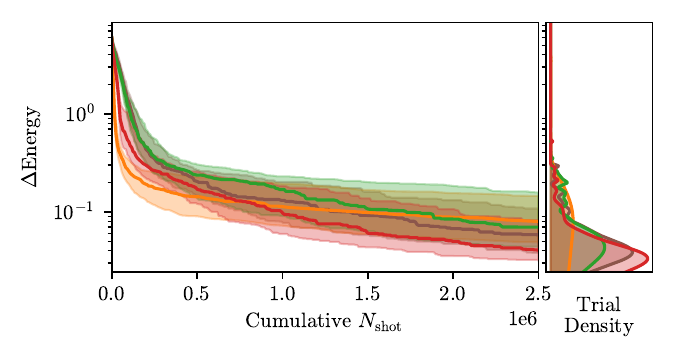}
    \includegraphics[width=0.49\textwidth]{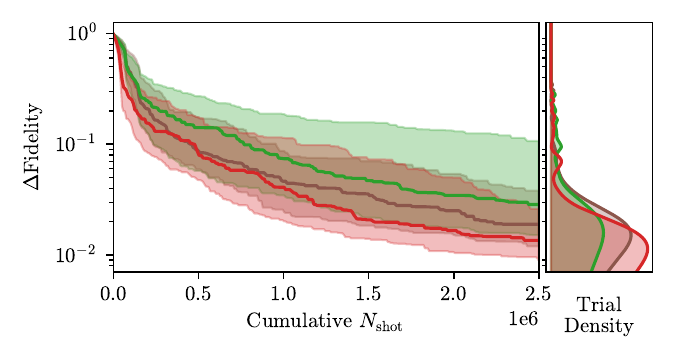}
    \centering\vskip -2ex
    \vskip -1ex
    \caption{Energy (left) and fidelity (right), as the difference from the ground-truth (see Eqs.~\eqref{eq:DeltaEnergy} and \eqref{eq:DeltaFidelity}),  achieved by our \methodname{}\added{-Center}{} and the baselines, NFT, SGLBO, and EMICoRe for the Ising model with $(Q, L) = (5,3)$. 
    Both plots are on a logarithmic scale, and the horizontal axis indicates the cumulative number of shots (per operator group) as the total quantum computation cost. The fidelity by SGLBO is not shown since the original code does not store the optimal parameters $\widehat{\bfx}$ to reproduce the quantum state $\vert\psi_{\widehat{\bfx}}\rangle$. However, given its slower convergence in terms of energy (left), we expect that the achieved fidelity is also worse than our \methodname{}\added{-Center}{}.}
    \label{fig:EnergyComparison}
\end{figure*}

\methodname{}-Bound ignores the existence of previous training data $\bfX$, and therefore gives a loose bound. 
 On the other hand, \methodname{}\added{-Center}{} makes use of the fact that GP already has low uncertainty ($\leq \kappa_{t-1}^2$) at the current best point $\widehat{\bfx}^{t-1}$ after the previous iteration, and adjust the observation accuracy $\breve{\sigma}_0^2$ at $\widehat{\bfx}^{t-1}$ and the accuracy $\breve{\sigma}^2$ at the other points separately. 
 Therefore, \methodname{}\added{-Center}{} is more efficient than \methodname{}-Bound.
 The difference in efficiency is expected to be more prominent in the converging phase, where the best point moves only slightly in each iteration, and therefore, the GP can already possess a low uncertainty over the subspace to be updated before new observations are acquired.

Once we specify the necessary observation noise $\breve{\bfsigma}$, the number of shots is chosen such that $\sigma^{*2}(N_{\mathrm{shot}}) = \breve{\sigma}_w^2$ for each observed point. 
Below, we describe the whole procedure of \methodname{}\added{-Center}, which equips NFT with the adaptive number of shots control. \added{For further details on the algorithm we defer the reader to \cref{sec:A.Pseudocode}. Throughout the manuscript, \methodname{} will always implicitly refer to \methodname{}-Center unless specified otherwise.
}

\begin{figure}[t]
    \centering
    \includegraphics[height=5ex]{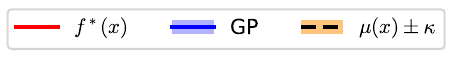}\\
    \includegraphics[width=0.49\textwidth]{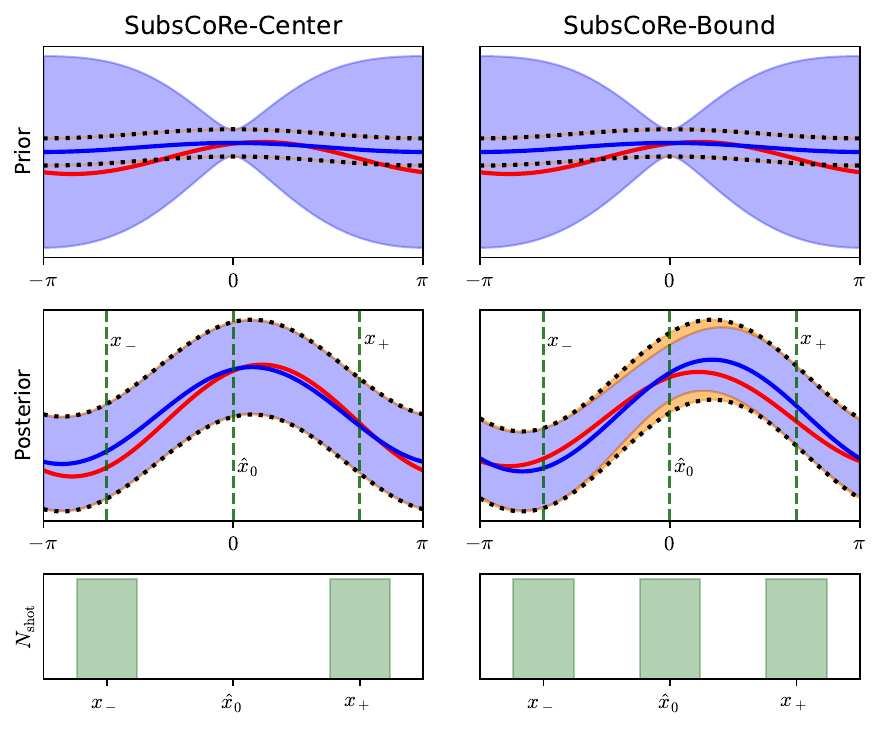}
    \vskip -1ex
    \caption{Prior (top) and posterior (middle) GP with  \methodname{}\added{-Center} (left) and \methodname{}-Bound (right) in an SMO step. 
    The red solid lines are the true function $f^*(\bfx)$, and the blue solid lines and shadows are the GP mean and uncertainty, respectively. The vertical lines represent the previous best point $\widehat{x}_0$ and the equidistant shifts ($x_{\pm}$) as visualized in the cartoon in~\cref{fig:cartoon_SubsCoRe}. The bottom row reports the number of shots required by \methodname{}\added{-Center} and \methodname{}-Bound at each observed point.
    Before the new observations (i.e., prior), 
    the previous best point $\widehat{x}_0$ is already in the CoRe, i.e., the uncertainty of GP is within the CoRe requirement (shown as black dashed lines).  After the new observations (i.e., posterior), the entire subspace is in the CoRe. We see the gap (highlighted in orange) between the CoRe requirement and the posterior uncertainty with \methodname{}-Bound.  This is because \methodname{}-Bound relies on a looser bound and assigns many shots to the previous best point $\widehat{x}_0$, on which the prior GP was already trained accurately (see bottom-right plot). 
    }
    \label{fig:GPEvolution}
    \vspace{-5mm}
\end{figure}

\paragraph{\methodname{} Algorithm}
\methodname{} is initialized with a random point $\bfX^0 = (\widehat{\bfx}^{0})$ and a first observation $\bfy^0 = \widehat{y}^0 = f^*(\widehat{\bfx}^0) + \varepsilon_0$ with observation noise $\breve{\sigma}^2 = \kappa_0^2$.  Then, it iterates the following procedure: in each iteration step $t$, 
\begin{enumerate}
\itemsep0em 
    \item Select an axis $d \in \{1, \ldots, D\}$ sequentially and observe the objective $\breve{\bfy} \in \mathbb{R}^{1 + 2V_d}$ at the $1 + 2V_d$ equidistant points 
$\breve{\bfX} =   \{ \widehat{\bfx}^{t-1} +  \alpha_w  \bfe_{{d}} ; \alpha_w = \frac{2w}{1 + 2V_d } \pi  \}_{w=0}^{2V_d}$ along the $d$-th axis.  Here, the observation variances $\breve{\bfsigma}$ are determined by \methodname{}\added{-Center} (or \methodname{}-Bound), using the CoRe threshold $\kappa_t^2$, as well as the training data $\bfX^{t-1}, \bfsigma^{t-1}$ up to the previous iteration.

    \item Train GP with updated training data $\bfX^t = (\bfX^{t-1}, \breve{\bfX}), \bfsigma^t = (\bfsigma^{t-1}, \breve{\bfsigma}), \bfy^t = (\bfy^{t-1}, \breve{\bfy})$.

    \item {Analytically find the minimum of the GP mean $$\widehat{\bfx}^t = \argmin_{\bfx'} \mu_{[\bfX^t, \bfsigma^t, \bfy^t]}(\bfx')$$ in the 1D subspace $\bfx' = \widehat{\bfx}^{t-1} + \alpha' \bfe_d$ by applying the trigonometric polynomial regression to the GP predictions at $1 + 2V_d$ equidistant points.}
    
    \item Update the best score by $ \widehat{y}^t =  \mu_{[\bfX^t, \bfsigma^t, \bfy^t]}(\widehat{\bfx}^t)$.

    \item Update the CoRe threshold for the next iteration by
\begin{align}
\! 
    \kappa^{t+1} & =\textstyle \max (C_0 ,\, -C_1\cdot
    \mathrm{Slope} (\{\widehat{y}^{t'}\}_{t' = t - T_{\mathrm{Ave}}}^t)).
    \label{eq:CoreThreshUpdate}
\end{align}
\end{enumerate}
In Eq.~\eqref{eq:CoreThreshUpdate}, $\mathrm{Slope}(\cdot)$ estimates the progress of optimization by linear regression to the $T_{\mathrm{Ave}}$ recent best values,
and $C_0, C_1 \geq 0$ are the hyperparameters controlling the lower bound and linear relation between $\kappa$ and the slope, respectively. \added{In our experiments we set $C_0$ so that the number of shots is at most 1024 per data point (and per operator group), which is a typical value achievable on current quantum hardwares. Upper-bounding the number of shots 
allows us to keep the observation cost under control in the later stage of optimization, where the energy improvement, and thus the threshold $\kappa$, tends to be small. We set $C_1 = 1$.
}

\begin{figure*}[t]
    \centering
    \includegraphics[height=4.5ex]{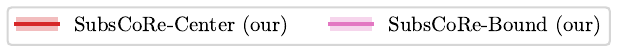}\\
    \includegraphics[width=0.49\textwidth]{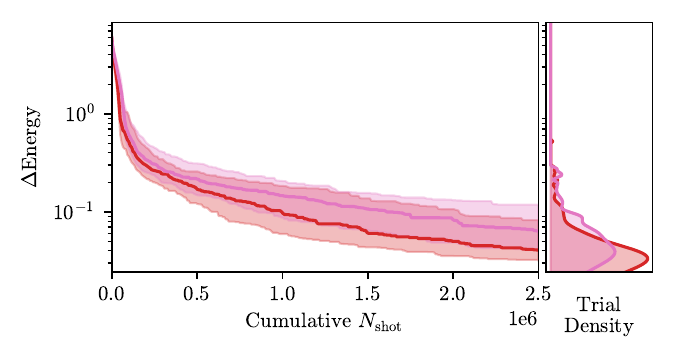}
    \includegraphics[width=0.49\textwidth]{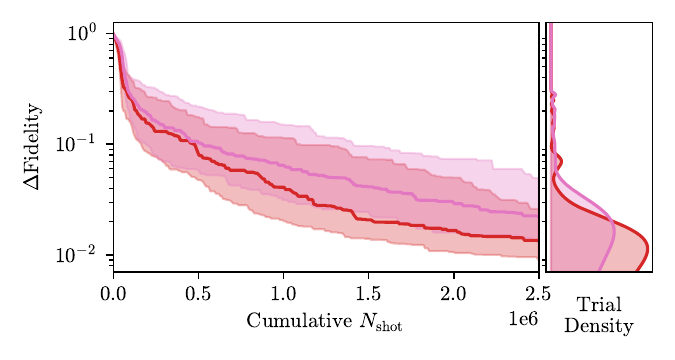}
    \centering
    \vskip -1ex
    \caption{%
    Energy (left) and fidelity (right), as the difference from the ground-truth (see Eqs.~\eqref{eq:DeltaEnergy} and~\eqref{eq:DeltaFidelity}), achieved by \methodname{}\added{-Center} and \methodname{}-Bound. 
     The inferior efficiency of \methodname{}-Bound is due to the fact that it relies on a looser bound.     
    }
\label{fig:AveragingEffect}
\end{figure*}

\section{Experiment}
\label{sec:Experiment}

\subsection{Setup}
We 
demonstrate the performance of our adaptive cost control method,
following the same experimental setup as in \citet{NEURIPS:Nicoli+:2023}.
Our Python implementation based on Qiskit~\cite{Abraham2019}, one of the most widely used library for developing and simulating quantum programs, \added{is available at \hyperlink{https://github.com/angler-vqe/subscore}{https://github.com/angler-vqe/subscore}}.

\paragraph{Hamiltonian and Quantum Circuit}
We focus on the quantum Heisenberg Hamiltonian with open boundary conditions,
\begin{align}
H =\textstyle -\sum_{i\in\{X,Y,Z\}}\left[\sum_{j=1}^{Q-1} (J_i \sigma_j^i \sigma_{j+1}^i) + \sum_{j=1}^{Q}h_i\sigma_j^{i}\right] ,
\label{eq:HeisenbergHamiltonian}
\end{align}
where $\{\sigma_j^i\}_{i\in\{X,Y,Z\}}$ are the Pauli operators acting on the $j$-th qubit. 
The Heisenberg Hamiltonian is a widely-studied benchmark for assessing VQE performance~\cite{tilly2022variational}
because it is of high practical relevance---for example, many lattice field theories can be represented as generalized spin chains~\cite{DiMeglio:2023nsa,Funcke:2023jbq}
---and the true ground-state wave function can be analytically computed for a small number $Q$ of qubits.   
For the quantum circuit, we adopt the commonly used $L$-layered \verb|Efficient SU(2)| circuit with open boundary conditions (see \citet{NEURIPS:Nicoli+:2023} for more details).

\paragraph{Evaluation Metrics}
We evaluate our methods with two metrics: the best achieved \textit{true energy} $f^*(\widehat{\bfx}) $ for $f^*(\cdot)$ defined in Eq.~\eqref{eq:VQEObjective}, and the \textit{fidelity} $\langle{\psi_{\mathrm{GS}}} \vert{\psi_{\widehat{\bfx}}}\rangle \in [0,1]$, which is the inner product between the true ground-state wave function $\vert{\psi_{\mathrm{GS}}}\rangle$ and the wave function $\vert{\psi_{\widehat{\bfx}}}\rangle$ corresponding to the optimized parameters $\widehat{\bfx}$.
For both metrics, we plot the difference (the smaller is the better) from the ground-truth optimum, i.e., 
\begin{align}
\Delta\mathrm{Energy} &= f^*(\widehat{\bfx}) - \langle{\psi_{\mathrm{GS}}}\vert  H  \vert{\psi_{\mathrm{GS}}}\rangle,
\label{eq:DeltaEnergy}\\
\Delta \mathrm{Fidelity}
&= \langle{\psi_{\mathrm{GS}}} \vert{\psi_{\mathrm{GS}}}\rangle
-
\langle{\psi_{\mathrm{GS}}} \vert{\psi_{\widehat{\bfx}}}\rangle
\notag\\
&= 1
-
\langle{\psi_{\mathrm{GS}}} \vert{\psi_{\widehat{\bfx}}}\rangle,
\label{eq:DeltaFidelity}
\end{align}
in the log scale.
Here, $\vert{\psi_{\mathrm{GS}}}\rangle$ and $\langle{\psi_{\mathrm{GS}}}\vert  H  \vert{\psi_{\mathrm{GS}}}\rangle$ are the 
ground-state wave function and the true ground-state energy, respectively, both of which are computed analytically.
As the quantum computation cost, we count the total number of measurement shots \emph{per operator group} (see the first footnote in \Cref{sec:B.VQE}), i.e., the cumulative sum of $N_{\mathrm{shot}}$ over all observations in the whole optimization process.

\paragraph{Baseline Methods}
We compare our \methodname{} approach to the state-of-the-art baselines, NFT \citep{nakanishi20} (with $\alpha = \pm \frac{2\pi}{3}$ shifts),
SGLBO \citep{SGLBO2022}, and EMICoRe \citep{NEURIPS:Nicoli+:2023}.
SGLBO is a gradient-based method equipped with adaptive shot control,
while NFT and EMICoRe are SMO-based approaches for which no adaptive shot control technique was available before our approach.  Therefore, we evaluate NFT and EMICoRe with a fixed number of shots $N_{\mathrm{shot}} = 1024$, i.e., the same setting as in \citet{NEURIPS:Nicoli+:2023}. 

Further details on the algorithms, including the parameter setting, and the experiments are given in Appendices 
\ref{sec:A.AlgorithmDetail} and \ref{sec:A.ExperimentDetail}, respectively.

\subsection{Performance Evaluation}
\label{sec:PerformanceEvaluation}
We evaluate our \methodname{} and the baseline methods on the Ising Hamiltonian, a special case of the Heisenberg Hamiltonian \eqref{eq:HeisenbergHamiltonian}  with the coupling parameters set to $J_X=-1,\, J_Y=J_Z=0, h_X=h_Y=0,\, h_Z=-1$, with the $(L=3)$-layered quantum circuit with $(Q=5)$-qubits.
\Cref{fig:EnergyComparison} shows the 
achieved energy \eqref{eq:DeltaEnergy}
and the fidelity \eqref{eq:DeltaFidelity}---as the difference from the ground truth---with the cumulative number of measurement shots in the horizontal axis.
\added{
On the right-hand side of each plot, we show 
the trial density
after $\totalshots{}$ measurement shots are used. We applied the kernel density estimation over the 100 independent seeded trials to produce the trial density.}
We observe that \methodname{}-Center outperforms the baselines
with statistical significance ($p$-value $<0.05$ according to the Wilcoxon signed-rank test), thus proving the usefulness of our adaptive cost control for SMO. 
\added{
We show the evolution of the number of measurement shots taken by \methodname{}-Center
as a function of the SMO steps
in \cref{sec:A.readout}.}


Figure \ref{fig:GPEvolution} shows the trained GP with 
\methodname{}-Center (left) and \methodname{}-Bound (right), prior (top) and posterior (middle) to the acquisition of the new observations for an SMO step.  We see that the posterior uncertainty with \methodname{}-Center is almost uniform and just inside the CoRe requirement, i.e.,  $\mu(\bfx) \pm \kappa$. Crucially, fine-tuning the number of measurement shots (bottom) allows not to consume shots at the previous best point $\widehat{x}_0$, which was already in the CoRe 
for the prior GP (see the top plot).
On the other hand, \methodname{}-Bound, which relies on a looser bound, allocates shots also to the previous best point, making some regions too accurate (highlighted by orange).  This explains the inferior efficiency of \methodname{}-Bound, which will be shown in \Cref{fig:AveragingEffect} in \Cref{sec:Discussion}.

\subsection{Discussion}
\label{sec:Discussion}

\paragraph{GP vs.\ Suffix Averaging}

The SGLBO baseline is not only equipped with adaptive shot control but also with \emph{suffix averaging} (SA), i.e., it takes the average of the best point trail over the last, e.g., 10\% iterations, i.e., $\widehat{\bfx} =\frac{1}{0.1 T} \sum_{t= 0.9T} ^T \widehat{\bfx}^t$.  
The idea behind SA is that, in the converging phase, the best point can fluctuate around the optimum because of the noisy observations, and therefore, taking the average over the optimization steps can better align the best point to the true optimum. Significant improvement by using SA was reported in 
\citet{SGLBO2022}. We argue that our \methodname{} enjoys this effect in a statistically more pronounced way as it uses the GP surrogate trained not only on the new observations but also on those from the previous steps. 
We emphasize that, while SGLBO also uses a GP surrogate, it is not trained on previous observations, and therefore, previous observations are only leveraged through the SA heuristics.

\paragraph{\methodname{}-Center vs.\ \methodname{}-Bound}

As discussed in \Cref{sec:PerformanceEvaluation},
the \methodname{}-Bound variant  does not consider the contribution from the previous observations.  A consequence can be observed in \Cref{fig:AveragingEffect}, 
where \methodname{}-Bound is outperformed by \methodname{}-Center, and the difference increases in the later phase of the optimization.

\section{Conclusion}
\label{sec:Conclustion}


The intrinsic probabilistic nature of measurements is one of the major drawbacks of quantum algorithms, to which noise-resilient statistical techniques are expected to make significant contributions.  Our \emph{subspace in confident region} (\methodname{}) approach for
 variational quantum eigensolvers (VQEs)
 has shown to improve the optimization efficiency by dynamically controlling the observation costs, 
 and thus reducing the quantum computation budget  without sacrificing the accuracy of the optimization.  Here, the \emph{uncertainty estimation}, a key implement in Bayesian statistics, plays an essential role.   
 Specifically, harnessing the notion of the confident region (CoRe) of Gaussian processes (GP), \methodname{} minimizes the observation costs while guaranteeing the required accuracy for prediction.  Our theory, built upon the unique physical property of VQEs and the analytic form of GP regression, proved the uniform uncertainty of GP predictions trained on equidistant observations, leading to simple algorithms that approximate the optimal cost distribution. The developed \methodname{} outperforms the state-of-the-art methods with and without adaptive cost control.  Our analysis also clarified the relation between GP with the VQE kernel and the Fourier analysis, which may facilitate further development of efficient VQE algorithms. In future work, we plan to run experiments on real quantum devices, and thus benchmark our approaches in the presence of quantum hardware noise. Another future plan is to apply our methods to different Hamiltonians with high practical relevance in quantum chemistry and lattice field theory.

\section*{Impact Statement}

This paper presents work whose goal is to advance the field of Machine Learning and Quantum Computing. There are many potential societal consequences of our work, none of which we feel must be specifically highlighted here.

\section*{Acknowledgements}
\added{
The authors thank the reviewers for their constructive comments and discussion for improving the paper. 
This work was supported by the German Ministry for Education and Research (BMBF) under the grant BIFOLD24B, the Einstein Research Unit Quantum Project (ERU-2020-607), the European Union’s HORIZON MSCA Doctoral Networks programme project AQTIVATE (101072344), the Deutsche Forschungsgemeinschaft (DFG, German Research Foundation) as part of the CRC 1639 NuMeriQS – project no. 511713970, under Germany’s Excellence Strategy -- Cluster of Excellence Matter and Light for Quantum Computing (ML4Q) EXC 2004/1 -- 390534769,
the European Union’s Horizon Europe Framework Programme (HORIZON) under the ERA Chair scheme with grant agreement no.\ 101087126,
and the Ministry of Science, Research and Culture of the State of Brandenburg within the Centre for Quantum Technologies and Applications (CQTA).}
\bibliography{camera-ready-modified}
\bibliographystyle{icml2024}

\newpage
\appendix
\onecolumn

\added{\section{Limitations}
\label{sec:A.limitations}
The limitations of our approach are inherently related to those of VQEs. In fact, applications of the proposed method on current quantum devices are mostly obstructed by hardware limitations. In addition to the probabilistic nature of the measurement process, which we tackled in this paper, other types of noise resulting from imperfect isolation from the environment and limited experimental control still significantly affect current quantum hardware. In light of this, the \methodname{} approach we proposed in this paper in principle does not suffer from crucial limitations per se, but indeed will also be affected by the same bottlenecks of VQEs. Nonetheless, we envision that leveraging machine learning techniques and Bayesian statistics, in particular BO and GP regression in this context, is particularly promising as it might allow us to cope better with the inherent probabilistic nature of such hybrid quantum-classical algorithms.
}

\section{Theoretical Details}

\section{Proofs of \Cref{thrm:UniformUncertainty} and \Cref{thrm:UniformUncertaintyUpperBound}} 
\label{sec:A.Proofs}

We use basic properties of the discrete Fourier transform, derived by the \emph{root of unity}, i.e., $\xi_L = e^{2\pi i/L}$.
Since
\begin{align}
  \sum_{l=0}^{L-1} \xi_L^{lw} = \frac{1 - \xi_{L}^{wL}}{1 - \xi_{L}^w} =  0 \qquad \mbox{ for } \qquad  w = 1, \ldots, L-1,
  \notag
\end{align}
we have 
\begin{align}
  \sum_{v=0}^{2V_d} e^{vw \frac{2\pi i  }{1 + 2V_d}} =  0 \qquad \mbox{ for } \qquad w = 1, \ldots, 2 V_d.
  \notag
\end{align}
Comparing the real and imaginary parts, we have
\begin{align}
\sum_{v=0}^{2V_d} \cos \left( vw \frac{2\pi   }{1 + 2V_d}\right) 
& =  
\begin{cases}
1 +2V_d & \mbox{ for } \qquad w=0, \\
0 & \mbox{ for } \qquad w = 1, \ldots, 2 V_d,
\end{cases}
\label{eq:BasicCos}\\
\sum_{v=0}^{2V_d} \sin \left( vw \frac{2\pi   }{1 + 2V_d}\right) 
& =  
0 \qquad \mbox{ for } \qquad w = 0, \ldots, 2 V_d.
\label{eq:BasicSin}
\end{align}
From the symmetry, 
\begin{align}
\sum_{v=1}^{V_d} \cos \left( vw \frac{2\pi   }{1 + 2V_d}\right) 
= \sum_{v=V_d + 1}^{2V_d} \cos \left( vw \frac{2\pi   }{1 + 2V_d}\right) ,
\label{eq:BasicSymmetry}
\end{align}
and thus
\begin{align}
1 + 2 \sum_{v=1}^{V_d} \cos \left( vw \frac{2\pi   }{1 + 2V_d}\right) 
=  0 \qquad \mbox{ for } \qquad w = 1, \ldots, 2 V_d.
\label{eq:BasicKernelEntry}
\end{align}

By using the basic equations \eqref{eq:BasicCos}--\eqref{eq:BasicKernelEntry}, we will derive the predictive variance of GP in the following.

The kernel matrix for the training points $\{\widehat{\bfx}  + \frac{2w}{1 + 2V_d } \pi \bfe_d\}_{w = 0}^{2V_d}$ is Toeplitz as
\begin{align}
{\bfK} 
&= 
\sigma_0^2
\begin{pmatrix}
\tau_0 & \tau_1  & \tau_2 &  \cdots & \tau_{2V_d-1}  & \tau_{2V_d}  \\
\tau_1 & \tau_0&\tau_1 &&& \tau_{2V_d-1} \\
\tau_2 & \tau_1 & \tau_0\\
\vdots & & & \ddots \\
  \tau_{2V_d-1} & & & & \tau_0 & \tau_1\\
 \tau_{2V_d} &  \tau_{2V_d-1} & & & \tau_1 & \tau_0
     \end{pmatrix}
     \in \mathbb{R}^{(1 + 2V_d) \times (1 + 2V_d) },
\notag
\end{align}
where
\begin{align}
\tau_w
&= 
\frac{  \gamma^{2} + 2\sum_{v=1}^{V_d}   \cos \left( \frac{2v w}{1 + 2 V_d}\pi  \right)}{ \gamma^{2} +  2V_d }.
\notag
\end{align}
The test kernel components for a test point at $\bfx' = \widehat{\bfx}  +\alpha' \bfe_d$ for arbitrary $\alpha' \in [0, 2\pi)$ are
\begin{align}
{\bfk}' 
&= 
\sigma_0^2
\begin{pmatrix}
\rho_0  \\
\rho_1 \\ 
\vdots \\
\rho_{2V_d}
\end{pmatrix},
\notag\\
{k}'' &= 
\sigma_0^2,
\notag
\end{align} 
where
\begin{align}
\rho_w
&=
  \frac{  \gamma^{2} + 2\sum_{v=1}^{V_d}   \cos \left( v\left(\frac{2w\pi}{1 + 2V_d} - \alpha'\right)  \right)}{ \gamma^{2} +  2V_d}.
  \notag
\end{align}

Eq.~\eqref{eq:BasicKernelEntry} implies that
\begin{align}
\tau_w
&= 
\begin{cases}
1 & \mbox{ for } \qquad w= 0,\\
 \frac{\gamma^2 -1}{\gamma^2 + 2V_d} & \mbox{ for } \qquad w= 1, \ldots, 2V_d,
\end{cases}
\notag
\end{align}
and therefore
\begin{align}
{\bfK} 
&= 
\sigma_0^2
\left(   \frac{1 + 2 V_d}{\gamma^2 + 2V_d} \bfI_{1 + 2V_d} + \frac{\gamma^2 -1}{\gamma^2 + 2V_d} \bfone_{1 + 2V_d} \bfone_{1 + 2V_d}^\T\right),
\notag
\end{align}
where $\bfI_N$ and $\bfone_N$ denote the $N$-dimensional identity matrix and the $N$-dimensional vector with all entries equal to one, respectively.
The matrix inversion lemma gives
\begin{align}
 & \hspace{-10mm} \left( \bfK + \sigma^{2}\bfI_{1 + 2V_d} \right)^{-1}\notag\\
  & =\frac{1}{\sigma_0^2}  \left( \left(\sigma^{2} /\sigma_0^2 + \frac{1 + 2V_d }{\gamma^2 + 2V_d} \right) \bfI_{1 + 2V_d} + \frac{\gamma^2 -1}{\gamma^2 + 2V_d} \bfone_{1 + 2V_d} \bfone_{1 + 2V_d}^\T \right)^{-1}
  \notag\\
  & = \frac{1}{\sigma_0^2} \left(\sigma^{2} /\sigma_0^2 + \frac{1 +2V_d }{\gamma^2 + 2V_d} \right)^{-1}  \left(  \bfI_{1 + 2V_d} +  \frac{\gamma^2 -1}{( \gamma^2  + 2V_d) \sigma^{2}/\sigma_0^2+1 + 2 V_d}\bfone_{1 + 2V_d} \bfone_{1 + 2V_d}^\T \right)^{-1}
  \notag\\
  & = \frac{1}{\sigma_0^2} \left(\sigma^{2} /\sigma_0^2 + \frac{1 +2V_d }{\gamma^2 + 2V_d}  \right)^{-1}
  \notag\\
  & \hspace{8mm}
  \left(
    \bfI_{1 + 2V_d} -  \frac{\gamma^2 -1}{( \gamma^2  + 2V_d) \sigma^{2}/\sigma_0^2+1 + 2V_d} \bfone_{1 + 2V_d}    \left(1 +  \frac{ (1 + 2V_d)(\gamma^2 -1)}{( \gamma^2  + 2 V_d) \sigma^{2}/\sigma_0^2+1 + 2V_d} \right)^{-1}  \bfone_{1 + 2V_d}^\T 
    \right)
  \notag\\
  & = \frac{1}{\sigma_0^2} \frac{\gamma^2 + 2V_d}{(\gamma^2 + 2V_d) \sigma^{2}/\sigma_0^2 + 1 + 2 V_d }
  \left(
    \bfI_{1 + 2V_d} -      \left(\frac{( \gamma^2 + 2V_d)\sigma^{2}/\sigma_0^{2}  +(1+2V_d) \gamma^2 } {\gamma^2 -1}  \right)^{-1}  \bfone_{1 + 2V_d}  \bfone_{1 + 2V_d}^\T 
    \right)
  \notag\\
    &=
\frac{1}{\sigma_0^2} a( \bfI_{1 + 2V_d} +  b \bfone_{1+2V_d} \bfone_{1+2V_d}^\T),
 \label{eq:A.TrainKernelInverse}
\end{align}
where
\begin{align}
a &=  \frac{\gamma^2 + 2V_d}{(\gamma^2 + 2V_d) \sigma^{2}/\sigma_0^2 + 1 + 2 V_d },
\notag\\
b &= -         \frac {\gamma^2 -1} {( \gamma^2 + 2V_d)\sigma^{2}/\sigma_0^{2}  +(1 + 2V_d) \gamma^2 } .
\notag
\end{align}

By using Eqs.~\eqref{eq:BasicCos} and \eqref{eq:BasicSin}, we have
\begin{align}
\|\bfk'\|^2
&=\sigma_0^4 \sum_{w = 0}^{2V_d} \rho_w^2
\notag\\
&=\sigma_0^4 \sum_{w = 0}^{2V_d} \left( \frac{ \gamma^2 + 2\sum_{v=1}^{V_d}   \cos \left( v\left(\frac{2w\pi}{1 + 2V_d} - \alpha' \right)  \right)}{\gamma^2 +  2V_d }\right)^2
\notag\\
&=\frac{ \sigma_0^4}{(\gamma^2 +  2V_d)^{2} }\sum_{w = 0}^{2V_d} \left\{ \gamma^4 + 4 \gamma^2\sum_{v=1}^{V_d}   \cos \left( v\left(\frac{2w\pi}{1 + 2V_d} - \alpha' \right)  \right) 
\right. \notag\\
& \hspace{10mm} \left.+ 4\sum_{v=1}^{V_d}  \sum_{v'=1}^{V_d}   \cos \left( v\left(\frac{2w\pi}{1 + 2V_d} - \alpha' \right) \right) \cos \left( v' \left(\frac{2w\pi}{1 + 2V_d} - \alpha' \right) \right)  \right\}
\notag\\
&=\frac{ \sigma_0^4}{(\gamma^2 +  2V_d)^{2} }\sum_{w = 0}^{2V_d} \left\{ \gamma^4 + 4 \gamma^2\sum_{v=1}^{V_d}   \cos \left( v\left(\frac{2w\pi}{1 + 2V_d} - \alpha' \right)  \right) 
\right. \notag\\
& \hspace{10mm} \left.
+ 2\sum_{v=1}^{V_d}  \sum_{v'=1}^{V_d} \left(  \cos \left( (v + v')\left(\frac{2w\pi}{1 + 2V_d} - \alpha' \right) \right) +  \cos \left((v- v') \left(\frac{2w\pi}{1 + 2V_d} - \alpha' \right) \right)\right)  \right\}
\notag\\
&=\frac{ \sigma_0^4}{(\gamma^2 +  2V_d)^{2} }\sum_{w = 0}^{2V_d} \Bigg\{ \gamma^4  + 4 (\gamma^2-1)\sum_{v=1}^{V_d}   \cos \left( v\left(\frac{2w\pi}{1 + 2V_d} - \alpha' \right)  \right)
 \notag\\
& \hspace{10mm} 
+ 2\sum_{v=1}^{V_d}  \sum_{v'=0}^{V_d} \left(  \cos \left( (v + v')\left(\frac{2w\pi}{1 + 2V_d} - \alpha' \right) \right) +  \cos \left((v- v') \left(\frac{2w\pi}{1 + 2V_d} - \alpha' \right) \right)\right)  \Bigg\}
\notag\\
&=\frac{ \sigma_0^4}{(\gamma^2 +  2V_d)^{2} } \Bigg\{ \gamma^4(1 + 2V_d ) + 4 (\gamma^2-1)\sum_{v=1}^{V_d}  \sum_{w = 0}^{2V_d}  \cos \left( v\left(\frac{2w\pi}{1 + 2V_d} - \alpha' \right)  \right)
\notag\\
& \hspace{10mm}
+ 2\sum_{v=1}^{V_d}  \sum_{v'=0}^{V_d} \sum_{w = 0}^{2V_d} \left(  \cos \left( (v + v')\left(\frac{2w\pi}{1 + 2V_d} - \alpha' \right) \right) +  \cos \left((v- v') \left(\frac{2w\pi}{1 + 2V_d} - \alpha' \right) \right)\right)  \Bigg\}
\notag\\
&=\frac{ \sigma_0^4}{(\gamma^2 +  2V_d)^{2} } \Bigg\{ \gamma^4 (1+ 2V_d)    + 4 (\gamma^2-1)\sum_{v=1}^{V_d} \sum_{w = 0}^{2V_d}  \cos \frac{2w v \pi}{1 + 2V_d}  \cos \left( v\alpha' \right)\notag\\
& \hspace{10mm}
+ 2\sum_{v=1}^{V_d}  \sum_{v'=0}^{V_d} \sum_{w = 0}^{2V_d} \Bigg(  \cos \frac{2w(v + v')\pi}{1 + 2V_d}  \cos \left( (v + v')\alpha' \right) +  \sin \frac{2w(v + v')\pi}{1 + 2V_d}  \sin \left( (v + v')\alpha' \right) 
\notag\\
& \hspace{10mm}
+  \cos \frac{2w(v - v')\pi}{1 + 2V_d}  \cos \left( (v - v')\alpha' \right) +  \sin \frac{2w(v - v')\pi}{1 + 2V_d}  \sin \left( (v - v')\alpha' \right)
\Bigg)  \Bigg\}
\notag\\
&=\frac{ \sigma_0^4}{(\gamma^2 +  2V_d)^{2} } \left\{ \gamma^4(1 + 2V_d )
+ 2\sum_{v=1}^{V_d}   \left(   1 + 2V_d
\right)  \right\}
\notag\\
&=\frac{ \sigma_0^4}{(\gamma^2 +  2V_d)^{2} } \left\{ \gamma^4(1 + 2V_d) 
+ 2 V_d   \left(   1 + 2V_d
\right)  \right\}
\notag\\
&=\sigma_0^4 \frac{ (1 + 2 V_d)(\gamma^4 + 2V_d)}{(\gamma^2 +  2V_d)^2 } ,
 \label{eq:A.TestKernelSquare}
\end{align}
and
\begin{align}
\|\bfk'\|_1 = \bfk'^\T \bfone_{1+2V_d}
&=\sigma_0^2 \sum_{w = 0}^{2V_d}  \frac{ \gamma^2 + 2\sum_{v=1}^{V_d}   \cos \left( v\left(\frac{2w\pi}{1 + 2V_d} - \alpha' \right)  \right)}{\gamma^2 +  2V_d }
\notag\\
&=\sigma_0^2  \frac{ \gamma^2(1 + 2V_d) + 2 \sum_{v=1}^{V_d}  \sum_{w = 0}^{2V_d}  \cos \left( v\left(\frac{2w\pi}{1 + 2V_d} - \alpha' \right)  \right)}{\gamma^2 +  2V_d }
\notag\\
&=\sigma_0^2  \frac{ \gamma^2(1 + 2V_d) + 2\sum_{v=1}^{V_d}  \sum_{w = 0}^{2V_d}  \left( \cos  \frac{2wv\pi}{1 + 2V_d}  \cos  v \alpha'  +  \sin  \frac{2wv\pi}{1 + 2V_d}  \sin  v \alpha'  \right)}{\gamma^2 +  2V_d }
\notag\\
&=\sigma_0^2 \frac{ (1 + 2 V_d)\gamma^2}{\gamma^2 + 2 V_d}.
 \label{eq:A.TestKernelSum}
\end{align}

With Eqs.~\eqref{eq:A.TrainKernelInverse}--\eqref{eq:A.TestKernelSum}, the predicitive variance is computed as 
\begin{align}
s(\bfx', \bfx')
&=
 k^{''} -   \bfk'^{\T}  \left( \bfK + \sigma^{2}\bfI_{1+2V_d} \right)^{-1}  \bfk'
 \notag\\
&=
\sigma_0^2-   \bfk'^{\T}  \frac{1}{\sigma_0^2} a  \left(  \bfI_{1+2V_d} + b \bfone_{1+2V_d} \bfone_{1+2V_d}^\T \right) \bfk'
 \notag\\
&=
\sigma_0^2-  \frac{1}{\sigma_0^2} a  \left(  \|\bfk'\|^2 + b (\bfk'^\T  \bfone_{1+2V_d} )^2 \right) 
 \notag\\
&=
\sigma_0^2-  \frac{1}{\sigma_0^2} a  \left(  \sigma_0^4 \frac{(1 + 2 V_d) \left(\gamma^4 + 2 V_d\right)}{(\gamma^2 + 2V_d)^2} + b \sigma_0^4 \frac{(1+2V_d)^2   \gamma^4}{(\gamma^2 + 2V_d)^2} \right) 
 \notag\\
&=
\sigma_0^2   \frac{(\gamma^2 + 2V_d)^2 - a\left( (1+2V_d) \left(\gamma^4 + 2V_d\right)  +     \frac {-  (1+2V_d)^2     \gamma^4    (\gamma^2 -1)} {( \gamma^2 + 2V_d)\sigma^{2}/\sigma_0^{2}  +(1 + 2V_d) \gamma^2 }\right)}{(\gamma^2 + 2V_d)^2}
 \notag\\
&=
\sigma_0^2   \frac{(\gamma^2 + 2V_d)^2 -   \frac{\gamma^2 + 2V_d}{(\gamma^2 + 2V_d) \sigma^{2}/\sigma_0^2 + 1 + 2V_d }\left(       \frac {(1+2V_d)( \gamma^4 + 2V_d) ( \gamma^2 + 2V_d)\sigma^{2}/\sigma_0^{2}  + (1+2V_d)^2 \gamma^2   (  \gamma^2 + 2V_d)    } {( \gamma^2 + 2V_d)\sigma^{2}/\sigma_0^{2}  +(1+2V_d) \gamma^2 }\right)}{(\gamma^2 + 2V_d)^2}
 \notag\\
&=
\sigma^2 \left(   \frac{  (\gamma^2 + 2V_d)^2 \sigma^{2}/\sigma_0^2  + (1+2V_d)^2\gamma^2     }{\left((\gamma^2 + 2V_d) \sigma^{2}/\sigma_0^2 + (1+2V_d) \right)\left(( \gamma^2 + 2V_d)\sigma^{2}/\sigma_0^{2}  +(1+2V_d) \gamma^2 \right) }\right),
\label{eq:A.UniformUncertainty}
\end{align}
which proves \Cref{thrm:UniformUncertainty}.

Since $(\gamma^2 + 2V_d) \sigma^{2}/\sigma_0^2 > 0$, we have
\begin{align}
s(\bfx', \bfx')
< \sigma^2 \left(   \frac{  (\gamma^2 + 2V_d)^2 \sigma^{2}/\sigma_0^2  + (1+2V_d)^2\gamma^2     }{\left( 1+2V_d \right)\left(( \gamma^2 + 2V_d)\sigma^{2}/\sigma_0^{2}  +(1+2V_d) \gamma^2 \right) }\right)
= \sigma^2, 
\notag
\end{align}
which proves \Cref{thrm:UniformUncertaintyUpperBound}.
In addition, asymptotic expansion of Eq.\eqref{eq:A.UniformUncertainty} for the case when $ \sigma^2 / \sigma_0^2 \ll 1$ gives the following corollary.
\begin{corollary}   
\label{thrm:UniformUncertaintyAsymptotic}
If $\frac{\sigma^{2}}{\sigma_0^2} \ll 1$,
\begin{align}
s_{\bfX}(\bfx', \bfx') 
&=
\begin{cases}
    \sigma^2 \left(  1 + O(\frac{\sigma^{2}}{\sigma_0^2}) \right) & \mbox { for } \gamma^2 = \Theta(1), \\
    \sigma^2 \left(  \frac{2V_d}{1 + 2V_d} + O(\frac{\sigma^{2}}{\sigma_0^2}) \right) & \mbox { for } \gamma^2 \to +0, \\
    \sigma^2 \left(  \frac{1}{1 + 2V_d} + O(\frac{\sigma^{2}}{\sigma_0^2}) \right) & \mbox { for } \gamma^2 \to \infty.
\end{cases}
\notag
\end{align}
\end{corollary}
\Cref{thrm:UniformUncertaintyAsymptotic} implies that, for any $\gamma^2 \in (0, \infty)$, the asymptotic uncertainty has the same form.  Setting $\gamma^2 \to +0$ removes the constant term, and thus the degree of freedom is reduced by one.  Similarly $\gamma^2 \to \infty$ removes the $2V_d$ terms except the constant, and thus the degree of freedom is reduced by $2V_d$.

\subsection{Relation to Fourier Analysis}
\label{sec:A.RelationToFourier}

\subsubsection{Equivalence between GP regression and Bayesian linear regression}

In general, it is known that the GP regression model
\begin{align}
p(y_n | \bfx_n, f(\cdot)) &= \mcN_D(y_n; f(\bfx_n), \sigma_n^2),
\label{eq:A.GPR.Likelihood}\\
p(f) &= \mathrm{GP}(f; 0(\cdot), k(\cdot, \cdot)),
\label{eq:A.GPR.Prior}\\
\end{align}
is equivalent to the Bayesian linear regression model
\begin{align}
p(y_n | \bfx_n, \bfb) &= \mcN_D(y_n; f(\bfx_n; \bfb), \sigma_n^2),
\label{eq:A.BLR.Likelihood}\\
f(\bfx; \bfb) & = \bfb^\T \bfphi (\bfx),
\label{eq:A.BLR.Function}\\
p(\bfb) &= \mcN_{H} (\bfb; \bfzero, \bfI_H),
\label{eq:A.BLR.Prior}
\end{align}
where $\bfphi (\bfx) \in \mathbb{R}^H$ is the finite-dimensional input feature such that $k(\bfx, \bfx') = \bfphi (\bfx)^\T \bfphi (\bfx')$.
This can be shown as follows.

Let $\bfX = (\bfx_1, \ldots, \bfx_N) \in \mathbb{R}^{D \times N}, \bfy = (y_1, \ldots, y_N)^\T \in \mathbb{R}^N$ be the training data, and 
$\bfPhi = (\bfphi(\bfx_1), \ldots, \bfphi(\bfx_N))  \in \mathbb{R}^{H \times N}$ be the corresponding input features.
The posterior of the Bayesian liner regression model \eqref{eq:A.BLR.Likelihood}--\eqref{eq:A.BLR.Prior} is
\begin{align}
p(\bfb | \bfX, \bfy)
&\propto
p(\bfy | \bfX, \bfb) p(\bfb)
\notag \\
&\propto  \exp \left(- \frac{( \bfy -  \bfPhi^\T \bfb)^\T \bfS^{-1} ( \bfy -\bfPhi^\T \bfb)}{2 } - \frac{\bfb^\T \bfI_{H}^{-1}\bfb }{2 } \right)
\notag \\
&\propto  \exp \left( - \frac{ \left(\bfb - \bfmu_b \right)^\T \bfSigma_b^{-1} \left(\bfb- \bfmu_b \right)}{2 } \right),
\notag 
\end{align}
where
\begin{align}
\bfmu_{b}
&= \bfSigma_b \bfPhi  \bfS^{-1} \bfy,
\notag\\
\bfSigma_b
&= \left( \bfPhi \bfS^{-1}\bfPhi^\T + \bfI_H \right)^{-1}
=\bfI_H -  \bfPhi \left( \bfS +  \bfPhi^\T\bfPhi \right)^{-1}\bfPhi^\T .
\notag
\end{align}
Therefore, the posterior is
\begin{align}
p(\bfb | \bfX, \bfy) &= \mcN_{H} (\bfb; \bfmu_{b}, \bfSigma_b).
\notag
\end{align}

The noiseless predictions $\bff' = (f(\bfx_1'), \ldots, f(\bfx_M'))^\T \in \mathbb{R}^{M}  $ at the test points $\bfX' = (\bfx_1', \ldots, \bfx_M') \in \mathbb{R}^{D \times M}$ with the corresponding features  $\bfPhi' = (\bfphi(\bfx_1'), \ldots, \bfphi(\bfx_M')) \in \mathbb{R}^{H \times M}$  follow
\begin{align}
p(\bff' | \bfX', \bfX, \bfy) &= \mcN_{M} (\bfy'; \bfmu_{f'}, \bfSigma_{f'}),
\notag\\
\bfmu_{f'} 
&= \bfPhi'^\T \bfmu_{b}
\notag\\
&= \bfPhi'^\T  \bfSigma_b \bfPhi  \bfS^{-1} \bfy,
\notag\\
 &= \bfPhi'^\T \left( \bfI_H -  \bfPhi \left( \bfS +  \bfPhi^\T\bfPhi \right)^{-1}\bfPhi^\T \right)\bfPhi  \bfS^{-1} \bfy
\notag\\
 &=\left(  \bfPhi'^\T \bfPhi -  \bfPhi'^\T \bfPhi \left( \bfS +  \bfPhi^\T\bfPhi \right)^{-1}\bfPhi^\T\bfPhi  \right) \bfS^{-1} \bfy
\notag\\
 &= \bfPhi'^\T \bfPhi \left( \bfI_N -  \left( \bfS +  \bfPhi^\T\bfPhi \right)^{-1}\bfPhi^\T\bfPhi  \right) \bfS^{-1} \bfy
\notag\\
 &= \bfK'^\T \left( \bfI_N -  \left( \bfS +  \bfK \right)^{-1}\bfK  \right) \bfS^{-1} \bfy
\notag\\
 &= \bfK'^\T \left( \bfS^{-1} -  \left( \bfS +  \bfK \right)^{-1}\bfK \bfS^{-1} \right)  \bfy
\notag\\
 &= \bfK'^\T \left( \bfS+ \bfK \right)^{-1}  \bfy
\label{eq:PosteriorMeanB}\\
\bfSigma_{f'} 
&=  \bfPhi'^\T  \bfSigma_b \bfPhi'
\notag\\
&=  \bfPhi'^\T  \left( \bfI_H -  \bfPhi \left( \bfS +  \bfPhi^\T\bfPhi \right)^{-1}\bfPhi^\T \right) \bfPhi'
\notag\\
&=   \bfPhi'^\T\bfPhi' -  \bfPhi'^\T\bfPhi \left( \bfS +  \bfPhi^\T\bfPhi \right)^{-1}\bfPhi^\T\bfPhi'
\notag\\
&=   \bfK'' -  \bfK'^\T \left( \bfS +  \bfK \right)^{-1}\bfK',
\label{eq:PosteriorCovarianceB}
\end{align}
which match the posterior \eqref{eq:GPPosterior}--\eqref{eq:GPPosteriorVar} of the GP regression model \eqref{eq:A.GPR.Likelihood}--\eqref{eq:A.GPR.Prior}.
Here, we obtained Eq.~\eqref{eq:PosteriorMeanB} by using the matrix inversion lemma
\begin{align}
(\bfA + \bfU \bfC \bfV)^{-1} = \bfA^{-1} -  \bfA^{-1}\bfU (\bfC^{-1} + \bfV \bfA^{-1} \bfU)^{-1} \bfV \bfA^{-1}
\notag
\end{align}
with $\bfA = \bfS, \bfU = \bfS, \bfC = \bfS^{-1}, \bfV = \bfK$, namely,
\begin{align}
(\bfS + \bfS \bfS^{-1} \bfK)^{-1} 
&= \bfS^{-1} -  \bfS^{-1}\bfS ((\bfS^{-1})^{-1} +\bfK \bfS^{-1} \bfS)^{-1} \bfK \bfS^{-1}
\notag\\
&= \bfS^{-1} -  (\bfS +\bfK )^{-1} \bfK \bfS^{-1}.
\notag
\end{align}

\subsubsection{Relation to Fourier Transform}

Similarly to the analysis in \Cref{sec:A.Proofs}, we compute the posterior mean of the GP regression at a test point $\bfx' = \widehat{\bfx}  +\alpha' \bfe_d$,
\begin{align}
\mu(\bfx')
&= \bfk'^{\T} \left(  \bfK + \sigma^2 \bfI_{1 + 2V_d} \right)^{-1} \bfy,
\notag\\
&=
\bfk'^{\T}  \frac{1}{\sigma_0^2} a  \left(  \bfI_{1 + 2V_d} + b \bfone_{1 + 2V_d} \bfone_{1 + 2V_d}^\T \right) \bfy
 \notag\\
&=
 \frac{1}{\sigma_0^2} a  \left(  \bfk'^{\T}\bfy + b \bfk'^\T  \bfone_{1 + 2V_d} \bfone_{1 + 2V_d}^\T \bfy \right) 
 \notag\\
&=
  \frac{1}{\sigma_0^2} a  \left( \sigma_0^2  \Bigg( \sum_{w = 0}^{2V_d} y_w \frac{ \gamma^2 + 2\sum_{v=1}^{V_d}   \cos \left( v\left(\frac{2w\pi}{1 + 2V_d} - \alpha' \right)  \right)}{\gamma^2 +  2V_d } \Bigg)
  + b \sigma_0^2 \frac{ (1 + 2 V_d)\gamma^2}{\gamma^2 + 2 V_d} \sum_{w = 0}^{2V_d} y_w \right) 
 \notag\\
&=
 \frac{a}{\gamma^2 + 2 V_d}    \sum_{w = 0}^{2V_d} y_w  \Bigg( \gamma^2 + 2\sum_{v=1}^{V_d}   \cos \left( v\left(\frac{2w\pi}{1 + 2V_d} - \alpha' \right)   \right) + b  (1 + 2 V_d)\gamma^2 \Bigg)  
 \notag\\
&=
 \frac{a}{\gamma^2 + 2 V_d}    \sum_{w = 0}^{2V_d} y_w  \Bigg( \gamma^2 + 2\sum_{v=1}^{V_d} \Bigg\{  \cos v\left(\frac{2w\pi}{1 + 2V_d} \right) \cos  v \alpha' +  \sin v\left(\frac{2w\pi}{1 + 2V_d} \right) \sin  v \alpha'  \Bigg\} 
 \notag\\
 & \hspace{30mm}
  -         \frac {(\gamma^2 -1)(1 + 2 V_d)\gamma^2} {( \gamma^2 + 2V_d)\sigma^{2}/\sigma_0^{2}  +(1 + 2V_d) \gamma^2 }   \Bigg) 
 \notag\\
&=
 \frac{a}{\gamma^2 + 2 V_d}    \sum_{w = 0}^{2V_d} y_w  \Bigg(  2\sum_{v=1}^{V_d} \Bigg\{  \cos v\left(\frac{2w\pi}{1 + 2V_d} \right) \cos  v \alpha' +  \sin v\left(\frac{2w\pi}{1 + 2V_d} \right) \sin  v \alpha'  \Bigg\} 
 \notag\\
 & \hspace{30mm}
  +         \frac {  \gamma^2 \left( ( \gamma^2 + 2V_d)\sigma^{2}/\sigma_0^{2}  +(1 + 2V_d) \gamma^2 \right)  - (\gamma^2 -1)  (1 + 2 V_d)\gamma^2} {( \gamma^2 + 2V_d)\sigma^{2}/\sigma_0^{2}  +(1 + 2V_d) \gamma^2 }  \Bigg) 
 \notag\\
&=
 \frac{1}{( \gamma^2 + 2V_d)\sigma^{2}/\sigma_0^{2}  + 1 + 2V_d }    \sum_{w = 0}^{2V_d} y_w  \Bigg(  2\sum_{v=1}^{V_d} \Bigg\{  \cos v\left(\frac{2w\pi}{1 + 2V_d} \right) \cos  v \alpha' +  \sin v\left(\frac{2w\pi}{1 + 2V_d} \right) \sin  v \alpha'  \Bigg\} 
 \notag\\
 & \hspace{30mm}
  +         \frac {  \gamma^2 ( \gamma^2 + 2V_d)\sigma^{2}/\sigma_0^{2}   + (1 + 2 V_d)\gamma^2} {( \gamma^2 + 2V_d)\sigma^{2}/\sigma_0^{2}  +(1 + 2V_d) \gamma^2 }   \Bigg) 
 \notag\\
&= \eta_0 +\sum_{v=1}^{V_d}  \left( \eta_{\cos, v} \sqrt{2}  \cos  v \alpha' +  \eta_{\sin, v} \sqrt{2}  \sin  v \alpha' \right),
 \notag
\end{align}
where
\begin{align}
\eta_{0}
&=    \frac {  \gamma^2 } 
{ ( \gamma^2 + 2V_d)\sigma^{2}/\sigma_0^{2}  +(1 + 2V_d) \gamma^2 }  
 \sum_{w = 0}^{2V_d} y_w,
\notag\\
\eta_{\cos, v}
&= \frac{1}{( \gamma^2 + 2V_d)\sigma^{2}/\sigma_0^{2}  + 1 + 2V_d }    \sum_{w = 0}^{2V_d}  y_w \sqrt{2} \cos v\left(\frac{2w\pi}{1 + 2V_d} \right)  ,
\notag\\
\eta_{\sin, v}
&= \frac{1}{( \gamma^2 + 2V_d)\sigma^{2}/\sigma_0^{2}  + 1 + 2V_d }    \sum_{w = 0}^{2V_d}   y_w\sqrt{2} \sin v\left(\frac{2w\pi}{1 + 2V_d} \right) .
\notag
\end{align}

When $\gamma = 1$,  
\begin{align}
\eta_{0}
&=    \frac {  1 } 
{( 1+ 2V_d) \left(\sigma^{2}/\sigma_0^{2}  + 1 \right) } 
 \sum_{w = 0}^{2V_d} y_w,
\notag\\
\eta_{\cos, v}
&=  \frac {  1 } 
{( 1+ 2V_d) \left(\sigma^{2}/\sigma_0^{2}  + 1 \right) } 
   \sum_{w = 0}^{2V_d}   y_w\sqrt{2} \cos v\left(\frac{2w\pi}{1 + 2V_d} \right)  ,
\notag\\
\eta_{\sin, v}
&=  \frac {  1 } 
{( 1+ 2V_d) \left(\sigma^{2}/\sigma_0^{2}  + 1 \right) } 
  \sum_{w = 0}^{2V_d}  y_w \sqrt{2} \sin v\left(\frac{2w\pi}{1 + 2V_d} \right)  ,
\notag
\end{align}
which is the regularized discrete Fourier transform, i.e., converges to the standard Fourier transform as $\sigma^{2}/\sigma_0^{2} \to 0$.

\begin{table}[t]
  \centering
  \caption{Algorithm-specific choice of parameters for EMICoRe and \methodname{} for all experiments (unless specified otherwise).}
  \vspace{0.3cm}
  \begin{tabular}{|l|c|c|}
    \toprule
    {} & \textbf{Algorithm-specific parameters} &  \\
    \midrule
    \midrule
    {\verb|--acq-params|} & \textbf{EMICoRe params} & as in~\citet{NEURIPS:Nicoli+:2023} \\
    \midrule
    \verb|func| & \verb|func=ei| & Base acq. func. type \\
    \verb|optim| & \verb|optim=emicore| & Optimizer type \\
    \verb|pairsize| ($J_{\mathrm{SG}}$) & \verb|20| & \# of candidate points \\
    \verb|gridsize| ($J_{\mathrm{OG}}$) & \verb|100| & \# of evaluation points \\
    \verb|corethresh| ($\kappa$) & \verb|1.0| & CoRe threshold $\kappa$ \\
    \verb|corethresh_width| ($T_{\mathrm{Ave}}$) & \verb|10| & \# averaging steps to update $\kappa$\\
    \verb|coremin_scale| ($C_0$) & \verb|0.0| & Coefficient $C_0$ for updating $\kappa$\\    \verb|corethresh_scale| ($C_1$) & \verb|1.0| & Coefficient $C_1$ for updating $\kappa$\\
    \verb|samplesize| ($N_{\mathrm{MC}}$) & \verb|100| & \# of MC samples \\
    \verb|smo-steps| ($T_{\mathrm{NFT}}$) & \verb|0| & \# of initial NFT steps \\
    \verb|smo-axis| & \verb|True| & Sequential direction choice\\
    \midrule
    \midrule
    {\verb|--acq-params|} & \textbf{\methodname{} params} & this paper\tablefootnote{All hyperparameters not specified in the table are 
    set to the default values in~\citet{NEURIPS:Nicoli+:2023}.} \\
    \midrule
    \verb|optim| & \verb|readout| & Optimizer type \\
    \verb|readout-strategy| & \verb|center/core| & Alg type \methodname{}/\methodname{}-Bound \\
    \verb|corethresh-strategy| & \verb|linreg| & Linear regression for $\kappa$ \\
    \verb|corethresh| ($\kappa$) & \verb|512| & Initial $N_{\textrm{shots}}$ for CoRe \\
    \verb|corethresh_width| ($T_{\mathrm{Ave}}$) & \verb|40| & \# averaging steps to update $\kappa$\\
    \verb|coremin_scale| ($C_0$) & \verb|1024| & Coefficient $C_0$ for updating $\kappa$\\    \verb|corethresh_scale| ($C_1$) & \verb|1.0| & Coefficient $C_1$ for updating $\kappa$\\
    \verb|coremetric| & \verb|readout| & Metric to set CoRe
    \\
    \bottomrule
  \end{tabular}\label{tab:algparams}
\end{table}

\section{Algorithm Details}
\label{sec:A.AlgorithmDetail}

\subsection{Parameter Setting}
\label{sec:A.ParameterSetting}
Every algorithm evaluated in our benchmarking analysis has several hyperparameters to be set.
For transparency and to allow the reproduction of our experiments, we detail the choice of parameters for EMICoRe and \methodname{} in~\cref{tab:algparams}. The SGLBO results were obtained using the original code from~\citet{SGLBO2022} and we used the default setting from the original paper. For the NFT run, we used the default parameters specified in~\cref{tab:defaultparams}. For 
(classical) computational efficiency,
all EMICoRe and \methodname{} runs used the inducer option retaining only the last $< 100$ measured points once more than 120 points were stored in the GP.  Specifically, after determining the points to be discarded, we train a temporary GP on those points and replace them with the pivot point right after the latest discarded points, along with its prediction by the temporary GP. This allows the GP to be of constant size throughout the optimization.

\added{\subsection{Pseudocode}
\label{sec:A.Pseudocode}
In this section we provide a detailed description of our proposed algorithm. Specifically, \cref{alg:SubscoreAlgFull} describes the full \methodname{} algorithm while \cref{alg:Subscore} focuses on the sub-routine used therein.}

\begin{center}
\begin{minipage}[t]{0.95\textwidth}
\vspace{0pt}
\begin{algorithm}[H]
  \footnotesize
  \SetKwInOut{Input}{Input}
  \SetKwInOut{Output}{Output}
  \SetKwInOut{Params}{Parameters}

  \Input{
    \begin{itemize}[nosep]
      \item $\hat{\bfx}^0$: initial starting point (best point)
      \item $\bar{N}_0$: initial number of shots for starting point
      \item $\kappa_0:$ Initial CoRe threshold at step $t=0$.
    \end{itemize}
  }
  \Params{
    \begin{itemize}[nosep]
      \item $V_d=1$
      \item $D:$ number of parameters to optimize, i.e., $\hat{\bfx}\in\mathbb{R}^D$.
      \item $\hat\alpha:$ shift from best point at the previous step (default to $\hat\alpha=\frac{2\pi}{3}$)
      \item $N_\text{tot-shots}:$ Total \# of shots, i.e., maximum allowed quantum computing budget.
    \end{itemize}
  }
  \Output{
    \begin{itemize}[nosep]
      \item $\hat{\bfx}_T:$ optimal choice of parameters for the quantum circuit.
    \end{itemize}
  }
  \BlankLine
  $n \gets 0$ \tcc{initialize consumed shot budget}
  $t \gets 0$ \tcc{initialize optimization step}
  $d \gets 0$ \tcc{initialize optimization subspace/ direction}
  \BlankLine
  $y_0 \gets $ \texttt{quantum\_circuit(parameters=}%
    $\hat{\bfx}^0$%
    \texttt{, shots=}%
    $\bar{N}_0$%
    \texttt{)}%
  \tcc{measure initial best point}
  $\bfX^{0},\, \bfy^{0},\, \bfsigma^{0} \gets
    (\hat{\bfx}_0),\,
    (y_0),\,
    (\frac{\eta^2}{N_0})
  $ %
  \tcc{initialize gaussian process}
  \BlankLine
  \While{$n < N_\text{tot-shots}$}{

    $\breve{\mathbf{X}} \gets \left(
      \hat{\bfx}^t - \hat\alpha\cdot\bfe_d,\, \hat{\bfx}^t,\, \hat{\bfx}^t + \hat\alpha\cdot\bfe_d
    \right)$ %
    \tcc{choose points to measure along d}

    $\bar{N}^{t+1} \gets
      \texttt{choose\_shots(}
      \bfX^{t},\, \bfy^{t},\, \bfsigma^{t},\, \breve{\mathbf{X}},\, \kappa_t,\, d \texttt{)}
    $ %
    \tcc{choose number of shots (Alg. \ref{alg:Subscore})}

    \BlankLine
    \For{$i \in \{1,\, ..., |\breve{\mathbf{X}}|\}$}{
      $\breve{y}_i \gets $ %
      \texttt{quantum\_circuit(parameters=}%
      $\breve{\mathbf{X}}_i$%
      \texttt{, shots=}%
      $\bar{N}_0$%
      \texttt{)}%
      \tcc{measure chosen points}

      $\breve{\sigma}_i \gets \frac{\eta^2}{\bar{N}^{t+1}_i}$
    }
    $\breve{\bfy}\,, \breve{\bfsigma} \gets %
      ( \breve{y}_1, ..., \breve{y}_{|\breve{\mathbf{X}}|})\,,
      ( \breve{\sigma}_1, ..., \breve{\sigma}_{|\breve{\mathbf{X}}|})
    $ \\

    \BlankLine
    $\bfX^{t+1},\, \bfy^{t+1},\, \bfsigma^{t+1} \gets
      (\bfX^{t}\,, \breve{\bfX}),\,
      (\bfy^{t}\,, \breve{\bfy}),\,
      (\bfsigma^{t}\,, \breve{\bfsigma})
    $ %
    \tcc{add new points to gaussian process}

    \BlankLine

    $\widehat{\bfx}^{t+1} \gets \argmin_{\bfx'} \mu_{|\bfX^{t+1}, \bfsigma^{t+1}, \bfy^{t+1}}(\bfx')$ %
    \tcc{find point with minimum GP mean on d}

    $\hat{y}^{t+1} \gets  \mu_{|\bfX^{t+1}, \bfsigma^{t+1}, \bfy^{t+1}}(\widehat{\bfx}^{t+1})$ %
    \tcc{remember current minimum GP mean as best value}

    \BlankLine

    \If{$t>T_{\mathrm{Ave}}$}{
      $ \kappa^{t+1} \gets \textstyle \max (C_0 ,\, -C_1\cdot \mathrm{Slope} (\{\widehat{y}^{t'}\}_{t' = (t+1) - T_{\mathrm{Ave}}}^{t+1})).$
      \tcc{update the CoRe threshold}
      \tcc{N.B. $\mathrm{Slope}(\cdot)$ estimates the progress of optimization by linear regression to the $T_{\mathrm{Ave}}$ recent best values.}
    }

    $t \gets t + 1$ \tcc{update the step}
    $n \gets n + \sum_i \bar{N}_i^{t+1}$ \tcc{update the consumed shot budget}
    $d \gets (d+1)\mod{D}$ \tcc{select the new subspace/direction to optimize}
  }
  \KwRet{
    $\hat{\bfx}_{{T}}$
  }
  \caption{
    \textbf{(SubsCoRe Algorithm)} Optimization scheme using the SubsCoRe
    subroutine, as described in \cref{alg:Subscore}, for finding the optimal parameters
    $\hat{\bf{x}}$ for the quantum circuit. 
    The optimization stops when the total number of
    measurement shots reaches the maximum number of observation shots allowed, i.e.,
    $N_{\mathrm{tot-shots}}$.
        To avoid cluttering notation, the algorithm is restricted to the case where $V_d=1$.  Generalization to an arbitrary $V_d$ is straightforward.
  }%
  \label{alg:SubscoreAlgFull}
\end{algorithm}
\end{minipage}
\end{center}

\begin{center}
\begin{minipage}[t]{0.95\textwidth}
\vspace{0pt}
\begin{algorithm}[H]
  \footnotesize

  \setcounter{AlgoLine}{0}
  \SetKwInOut{Input}{Input}
  \SetKwInOut{Output}{Output}
  \SetKwInOut{Params}{Parameters}

  \Input{
    \begin{itemize}[nosep]
      \item $\bfX^t, \bfy^t, \bfsigma^t:$ Gaussian Process at step $t$
      \item $\breve\bfX:$ Points for which to choose the number of shots (in order: left shift, center, right shift; see line 7 of Alg.~\ref{alg:SubscoreAlgFull})
      \item $\kappa:$ CoRe threshold at step $t$.
      \item $d:$ direction along which $\breve\bfX$ are distributed
    \end{itemize}
  }
  \Params{
    \begin{itemize}[nosep]
      \item $V_d=1$
      \item $\eta^2:$ measurement variance using a single shot.
    \end{itemize}
  }
  \Output{
    \begin{itemize}[nosep]
      \item $ \bar{N}_{t+1}:$ shots for performing new measurements at step $t+1$.
    \end{itemize}
  }

  \BlankLine
  \Begin{

    $\mcS_d(\breve{\bfX}_1) \gets \{\breve{\bfX}_1 + {\alpha}\cdot \bfe_d\,;\,\,\alpha \in [0, 2\pi)\};$
    \tcc{define points along d}

    \BlankLine

    $\breve\sigma_\pm \gets \kappa$ \tcc{initialize shifted point variance to worst case}

    \BlankLine

    \For{$\tilde{\sigma}\in [\eta, \kappa]$}{
      \tcc{create temporary GP copies, add chosen points with $\tilde\sigma$ observation noise}
      $\bfX^{t'},\, \bfy^{t'},\, \bfsigma^{t'} \gets
        (\bfX^{t}\,, \breve{\bfX}),\,
        (\bfy^{t}\,, 0, 0, 0),\,
        (\bfsigma^{t}\,, \tilde{\sigma}, \tilde{\sigma}, \tilde{\sigma})
      $ \\

      \tcc{get smallest observation noise for which temporary GP's prediction of all points along d are within the core}
      \If{
        $(\forall \bfx\in \mcS_d(\breve{\bfX}_1)) ~:~ s_{|\bfX^{t'}, \bfsigma^{t'}} (\bfx, \bfx) \leq \kappa^2)$
        $\land$
        $(\breve\sigma_\pm > \tilde\sigma)$
      }{
        $\breve\sigma_\pm \gets \tilde\sigma$
      }
    }
    $\bar{N}_\pm \gets \frac{\eta^2}{\breve\sigma_\pm}$
    \tcc{compute shots from variance through single shot variance $\eta^2$}

    \BlankLine
    \If{ using SubsCoRe-Bound }{
      $\bar{N}_{t+1} \gets (\bar{N}_\pm, \bar{N}_\pm, \bar{N}_\pm)$ \tcc{choose the same number of
      shots for each point}
      \KwRet{$\bar{N}_{t+1}$}
    } \tcc{otherwise, we use SubsCoRe-Center, where we reevaluate the center}

    \BlankLine

    $\breve\sigma_0 \gets \kappa$ \tcc{initialize center point variance to worst case}

    \BlankLine
    \For{$\tilde{\sigma}\in [\eta, \kappa]$}{
      \tcc{create temporary GP copies, add shifted points with $\breve\sigma_\pm$ and center with
      $\tilde\sigma$ observation noise}
      $\bfX^{t'},\, \bfy^{t'},\, \bfsigma^{t'} \gets
        (\bfX^{t}\,, \breve{\bfX}),\,
        (\bfy^{t}\,, 0, 0, 0),\,
        (\bfsigma^{t}\,, \breve\sigma_\pm, \tilde{\sigma}, \breve\sigma_\pm)
      $ \\

      \tcc{get smallest observation noise for which temporary GP's prediction of all points along d are within the core}
      \If{
        $(\forall \bfx\in \mcS_d(\breve{\bfX}_1)) ~:~ s_{|\bfX^{t'}, \bfsigma^{t'}} (\bfx, \bfx) \leq \kappa^2)$
        $\land$
        $(\breve\sigma_0 > \tilde\sigma)$
      }{
        $\breve\sigma_0 \gets \tilde\sigma$
      }
    }
    $\bar{N}_0 \gets \frac{\eta^2}{\breve\sigma_0}$
    \tcc{compute shots from variance through single shot variance $\eta^2$}

    \BlankLine

    $\bar{N}_{t+1} \gets (\bar{N}_\pm, \bar{N}_0, \bar{N}_\pm)$
    \tcc{choose shots for center different from shifted points}

    \KwRet{
      $\bar{N}_{t+1}$
    }
  }
  \caption{%
    \textbf{(SubsCoRe Subroutine to identify the number of shots)}
    Find the number of shots for the points chosen to be measured in SubsCoRe. Can be used for
    either SubsCoRe-Bound, or SubsCoRe-Center.
  }%
  \label{alg:Subscore}
\end{algorithm}
\end{minipage}
\end{center}

\section{Experimental Details}
\label{sec:A.ExperimentDetail}

\subsection{Experimental Setting}

As mentioned in the main text, we follow the experimental setup in~\citet{NEURIPS:Nicoli+:2023}. Specifically, starting from the quantum Heisenberg Hamiltonian, we focus on the special case of the Ising Hamiltonian at the critical point by choosing the suitable couplings, namely
$$
     \textrm{Ising Hamiltonian at criticality: } J=(-1.0,\,0.0,\,0.0);\,h=(0.0,\,0.0,\,-1.0).
$$
It is important to note that for the system at hand, this choice of parameters already represents a challenging setup for a fixed number of qubits, as discussed in Sec. I.2 in~\citet{NEURIPS:Nicoli+:2023}.

For all methods,
we stopped the optimization when a maximum number of cumulative shots (total measurement budget on the quantum computer) is reached; unless specified otherwise we set this cutoff to $N^{\textrm{max}}_{\textrm{shots}}=\totalshots{}$.
We based our implementation on the EMICoRe code available on GitHub~\citep{emicore_GH_2023} at \hyperlink{https://github.com/angler-vqe/emicore}{\textit{https://github.com/emicore/emicore}}. 
Since EMICoRe and \methodname{} use a GP combined with the VQE kernel~\cite{NEURIPS:Nicoli+:2023}, we need to set the kernel parameters $\gamma$ and $\sigma_0$. Similarly to \citet{NEURIPS:Nicoli+:2023}, we set $\sigma_0$ based on a rough approximation of the ground-state energy, and optimize $\gamma$ by the leave-one-out cross validation with grid-search (90 points) in the range $[\gamma_{\textrm{min}}, \gamma_{\textrm{max}}] = [\sqrt{2}, 20]$.
$\gamma_{\textrm{max}}$ 
and the frequency of the $\gamma$ update
can be specified by \verb|max_gamma| and \verb|interval|, respectively, in our code (see~\cref{tab:defaultparams}). 
For NFT, we used the fixed equidistant shifts $\alpha=\pm \frac{2\pi}{3}$. 

Each experiment shown in the paper has been repeated 100 times (independent seeded trials) with different starting points determined by the seeds. We aggregated the statistics from these independent seeded trials and presented them in our plots. For a fair comparison, we used the same starting point in each trial (i.e., for each corresponding seed) for all algorithms.
All experiments were conducted on Intel Xeon Silver 4316 @ 2.30GHz CPUs.
\begin{table}[t]
  \centering
  \caption{Default choice of circuit parameters and hyperparameter optimization (for EMICoRe and \methodname{}) in all experiments (unless specified otherwise).}
  \vspace{0.3cm}
  \begin{tabular}{|l|c|c|}
    \toprule
    {} & \textbf{Deafult params} &  \\
    \midrule
    \verb|--n-qbits| & \verb|5| &  \# of qubits \\
    \verb|--n-layers| & \verb|3| & \# of circuit layers \\
    \verb|--circuit| & \verb|esu2| & Circuit name \\
    \verb|--pbc| & \verb|False| & Open Boundary Conditions \\
    \verb|--n-iter| & \verb|3*10**6| & \# max number of readouts \\
    \verb|--iter-mode| & \verb|readout| & \# max number of readouts \\
    \verb|--kernel| & \verb|vqe| & Name of the kernel \\
    \midrule
    {\verb|--hyperopt|} & \textbf{Hyperparams optim} & \\
    \midrule
    \verb|optim| & \verb|grid| & Grid-search optimization of $\gamma$ \\
    \verb|max_gamma| & \verb|20| & Max value for $\gamma$ \\
    \verb|interval| & \verb|100*1+20*9+10*100| & Scheduling for grid-search \\
    \verb|steps| & \verb|90| & \# steps in grid \\
    \verb|loss| & \verb|loo| & Loss type
    \\
    \bottomrule
  \end{tabular}\label{tab:defaultparams}
\end{table}

\begin{figure}[t]
    \centering
    \includegraphics[height=5ex]{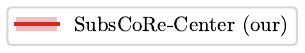}\\
    \includegraphics[width=0.45\columnwidth]{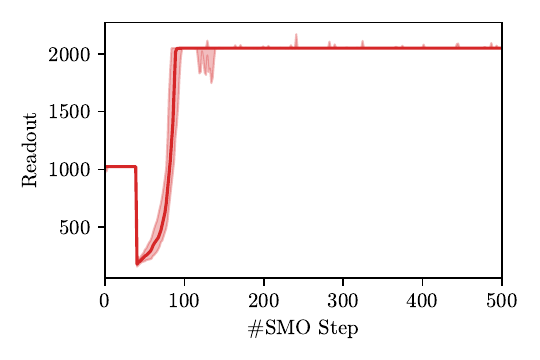}\\
    \vskip -1ex
    \caption{\added{Number of measurement shots (Readout) taken by \methodname{}-Center for each SMO step. The x-axis indicates the SMO step while the y-axis represents the total number of measurement shots taken at every step (i.e., distributed over the three observed points per step). 
    }}
    \label{fig:readout}
    \vspace{-5mm}
\end{figure}

\added{\subsection{Number of Shots per SMO Step}
\label{sec:A.readout}
The adaptive number of total shots (for all measurements combined) per SMO step for \methodname{}-Center (red) is shown in \cref{fig:readout}. 
}

\end{document}